\documentclass[journal]{IEEEtran}
\usepackage{courier}
\usepackage{amsmath,amsfonts}
\usepackage{algorithmic}
\usepackage{algorithm}
\usepackage{array}
\usepackage[caption=false,font=normalsize,labelfont=sf,textfont=sf]{subfig}

\usepackage{cancel}
\usepackage{textcomp}
\usepackage{stfloats}
\usepackage{url}
\usepackage{verbatim}
\usepackage{graphicx}
\usepackage{cite}
\usepackage{hyperref}
\usepackage{booktabs}  
\usepackage{multirow}  
\usepackage{acronym}
\usepackage{amssymb}
\usepackage{amsmath}
\usepackage{mathtools}
\usepackage{enumitem}
\usepackage{ifthen}
\usepackage[T1]{fontenc}
\usepackage[normalem]{ulem}
\usepackage[table, dvipsnames]{xcolor}
\definecolor{onsetColor}{RGB}{220,235,250} 
\definecolor{sampleColor}{RGB}{255,250,205} 
\definecolor{noneColor}{RGB}{240,240,240}  

\acrodef{CQT}{Constant-$Q$ Transform}
\acrodef{DDSP}{Differentiable Digital Signal Processing}
\acrodef{MIR}{Music Information Retrieval}
\acrodef{SSL}{Self-Supervised Learning}
\acrodef{TCN}{Temporal Convolutional Network}
\acrodef{VAD}{Voice Activity Detection}
\acrodef{GRU}{Gated Recurrent Unit}
\acrodef{MLP}{Multi-Layer Perceptron}
\acrodef{DNN}{Deep Neural Network}
\acrodef{CNN}{Convolutional Neural Network}
\acrodef{RNN}{Recurrent Neural Network}
\acrodef{DSS}{Drum Source Separation}
\acrodef{ADT}{Automatic Drum Transcription}
\acrodef{DTD}{Drum Transcription of Drum-only recordings}
\acrodef{NMF}{Non-negative Matrix Factorization}
\acrodef{NMFD}{Non-negative Matrix Factor Deconvolution}
\acrodef{MSS}{Music Source Separation}
\acrodef{OSS}{One-shot drum Sample Synthesis}
\acrodef{DSP}{Digital Signal Processing}
\acrodef{STFT}{Short-Time Fourier Transform}
\acrodef{FiLM}{Feature-wise Linear Modulation}
\acrodef{TF}{Time-Frequency}
\acrodef{SI-SDR}{Scale-Invariant Signal-to-Distortion Ratio}
\acrodef{nSDR}{Signal-to-Distortion Ratio}
\acrodef{LSD}{Log Spectral Distance}
\acrodef{PES}{Predicted Energy in Silence}
\acrodef{IDM}{Inverse Drum Machine}

\DeclareMathOperator*{\argmax}

\newboolean{devmode}
\setboolean{devmode}{false}

\ifthenelse{\boolean{devmode}}
{       
        \newcommand{\TODO}[1]{\textcolor{red}{TODO: #1}}
    \newcommand{\hash}[1]{}
    \newcommand{\bernardo}[1]{{\color{blue}bern: #1}}
    
    \newcommand{\gp}[1]{{\color{green}$\;^{GP:}$ \small{#1}}}
}{
    \newcommand{\TODO}[1]{}
        \newcommand{\hash}[1]{}
    \newcommand{\bernardo}[1]{}
    \newcommand{\gp}[1]{}

}

\newboolean{rebuttal}
\setboolean{rebuttal}{false}

\ifthenelse{\boolean{rebuttal}}
{
    \newcommand{\new}[1]{{\color{blue}#1}}
    
}{
    \newcommand{\new}[1]{#1}
    \renewcommand{\sout}[1]{}
    
}

\newboolean{usepng}
\setboolean{usepng}{true} 

\ifthenelse{\boolean{usepng}}{
    \DeclareGraphicsExtensions{.png,.pdf} 
}{
    \DeclareGraphicsExtensions{.pdf,.png} 
}

\newcommand{\nothing}[1]{}

\newcommand{\mathvar}[1]{\ensuremath{#1}}

\newcommand{\LL}[1]{\mathcal{L}_{\text{#1}}}

\newcommand{\sampleduration}{\mathvar{1}}
\newcommand{\duration}{\mathvar{6}}

\newcommand{\hopsizems}{\mathvar{11.6}}
\newcommand{\activationrate}{\mathvar{86.1}}

\newcommand{\numlayers}{\mathvar{12}}
\newcommand{\kernelsize}{\mathvar{15}}
\newcommand{\nparams}{\mathvar{542.1}K}
\newcommand{\onesecondsamples}{\mathvar{R}}
\newcommand{\samplerate}{\mathvar{44.1} kHz}

\newcommand{\framefeatures}{\mathvar    {\mathbf{z}}}

\newcommand{\frames}{\mathvar{M}}
\newcommand{\insts}{\mathvar{K}}

\newcommand{\activation}{\mathvar{\mathbf{a}}}

\newcommand{\velocity}{\mathvar{\mathbf{v}}}
\newcommand{\nembedding}{\mathvar{D_e}}
\newcommand{\nframefeatures}{\mathvar{D_z}}
\newcommand{\conditioning}{\mathvar{\mathbf{c}}}

\newcommand{\embedding}{\mathvar{\mathbf{e}}}
\newcommand{\sequencer}{\mathvar{\operatorname{Sequencer}}}
\newcommand{\gain}{\mathvar{\mathbf{g}}}
\newcommand{\samples}{\mathvar{\mathbf{w}}}
\newcommand{\lrecon}{\mathvar{\mathcal{L}_\text{recon}}}
\newcommand{\lemb}{\mathvar{\mathcal{L}_\text{emb}}}
\newcommand{\ltrans}{\mathvar{\mathcal{L}_\text{trans}}}
\newcommand{\oneshots}{\samples}
\newcommand{\onsets}{\mathvar{\mathbf{o}}}
\newcommand{\x}{\mathvar{\mathbf{x}}}

\newcommand{\hatxsynth}{\mathvar{\hat{\mathbf{x}}_\text{synth}}}

\newcommand{\hatxmask}{\mathvar{\hat{\mathbf{x}}_\text{mask}}}

\newcommand{\hatX}{\mathvar{\hat{\mathbf{X}}}}

\newcommand{\tracklength}{\mathvar{T}}
\newcommand{\larsnet}{\texttt{LarsNet}}
\newcommand{\larsnetmono}{\texttt{LarsNet Mono}}
\newcommand{\nmfd}{\texttt{NMFD}}
\newcommand{\nmfdthree}{\texttt{NMFD 3}}
\newcommand{\nmfdonea}{\texttt{NMFD 1A}}
\newcommand{\nmfdoneb}{\texttt{NMFD 1B}}
\newcommand{\ours}{\texttt{IDM}}
\newcommand{\oracle}{\texttt{Oracle}}
\newcommand{\oursonsets}{\texttt{\ours + onsets}}

\newcommand{\logmel}{Log-Mel}

\begin{document}

\title{The Inverse Drum Machine: Source Separation Through Joint Transcription and Analysis-by-Synthesis}

\author{
Bernardo Torres, Geoffroy Peeters, Gaël Richard 
\thanks{B. Torres, G. Peeters, and G. Richard are with the Laboratoire de Traitement et Communication de l’Information (LTCI), Télécom Paris, Institut Polytechnique de Paris, 91120 Palaiseau, France.}
}



\maketitle

\begin{abstract}

   We present the \acf{IDM}, a novel approach to \acl{DSS} that leverages an analysis-by-synthesis framework combined with deep learning. 
   Unlike recent supervised methods that require isolated stem recordings for \new{training}, our approach \new{is trained} on drum mixtures with only transcription annotations. 
   IDM integrates \acl{ADT} and \acl{OSS}, jointly optimizing these tasks in an end-to-end manner. By convolving synthesized one-shot samples with estimated onsets, akin to a drum machine, we reconstruct the individual drum stems and train a \acl{DNN} on the reconstruction of the mixture. 
   Experiments on the StemGMD dataset demonstrate that IDM achieves separation quality comparable to state-of-the-art supervised methods that require isolated stems data. \sout{, while significantly outperforming matrix decomposition baselines}

\end{abstract}





\IEEEPARstart{I}{n} Western popular music, the rhythmic foundation typically relies on percussion instruments from a standard drum kit comprising kick drum, snare drum, and hi-hat, while additional elements such as cymbals, tom-toms, and auxiliary percussions provide timbral complexity and rhythmic variation. 
Music producers and engineers often need to adjust individual drum instruments separately for remixing, rebalancing, effects processing, or creating educational materials \cite{fitzgeraldAutomaticDrumTranscription2004, dittmarRealtimeTranscriptionSeparation2014}. 
Ideally, music production would use isolated recordings of each drum instrument (known as "stems"), allowing for precise control during mixing. However, these instruments are usually played simultaneously and by the same performer, resulting in recordings in which all elements are mixed into a single audio stream. Obtaining these separated stems during recording requires multiple microphones (leading to microphone bleeding) or asking musicians to play in unnatural conditions \cite{gilletENSTDrumsExtensiveAudiovisual2006}. The need for tools that can extract individual drum stems from already mixed recordings has led to growing interest in \ac{DSS}\new{\cite{gilletTranscriptionSeparationDrum2008,dittmarReverseEngineeringAmen2016, mezzaDeepDrumSource2024}}.

In the professional music production industry, commercial tools such as DrumsSSX\footnote{\url{https://fuseaudiolabs.de/\#/pages/product?id=300867907}} and Fadr\footnote{\url{https://fadr.com/drum-stems}} offer drum stem separation capabilities. These solutions, however, are proprietary and still have limitations in separation quality and flexibility.

\ac{DSS} is challenging due to the acoustic properties of percussion sounds. Many percussive sounds lack a clear harmonic structure and exhibit broad spectral characteristics that \new{typically} overlap in both time and frequency domains. Additionally, there exists a wide diversity of percussive instruments, classified roughly into membranophones (e.g., kick drum, snare drum, toms) and idiophones (e.g., cymbals, hi-hats) \cite{rossingSciencePercussionInstruments2000}. While "drum", more strictly defined, refers to membranophones, we \new{follow common usage for simplicity and} use the term broadly to encompass all percussive instruments in this work. Each instrument can produce several sounds depending on the playing technique, further complicating the separation task. In addition to these complexities, drum-related music information research has received relatively limited attention, possibly reflecting Western music's historical emphasis on melodic and harmonic elements rather than rhythmic foundations found in traditions such as Indian tabla playing or African percussion-centered music.

In a broader context, \ac{MSS} has gained significant attention due to the impressive performance of deep learning methods. These approaches have achieved state-of-the-art results by training \acp{DNN} in a supervised manner using ground-truth separated stems \cite{hennequinSpleeterFastEfficient2020, defossezHybridSpectrogramWaveform2021, mezzaDeepDrumSource2024}. However, \ac{MSS} benchmarks typically focus on isolating just four classes: vocals, drums, bass, and other instruments, considering the drum section as a single stem. 

Recent works have capitalized on multitrack synthetic datasets generated by drum machines and programming software \cite{mezzaDeepDrumSource2024,mezzaBenchmarkingMusicDemixing2024} for training \ac{DSS} models. Simultaneously, the \ac{MSS} literature has shown promising results from joint separation and transcription approaches \cite{manilowSimultaneousSeparationTranscription2020a}. Complementing these developments, data-driven analysis-by-synthesis methods (\new{i.e., learning to decompose a signal by resynthesizing it from estimated parameters)} have demonstrated advances in \sout{reconstruction-based} techniques that do not require ground-truth data \cite{cheukEffectSpectrogramReconstruction2020, choiDeepUnsupervisedDrum2019, schulze-forsterUnsupervisedMusicSource2023}. 

In this work, we propose an analysis-by-synthesis, multitask approach to \ac{DSS}. By modeling drum tracks as sequences of one-shot samples triggered at precise times, similar to a drum machine, we train a \ac{DNN} to recover both the transcription and one-shot samples from the reconstruction of the mixture.
 \ac{DSS} is then achieved via audio synthesis from the "inverted" drum machine. Due to the strong inductive bias in our approach, we can bypass the need for isolated stems during training while achieving results comparable to state-of-the-art methods. \new{This trade-off is advantageous, as datasets with transcription annotations are generally more accessible than those with clean, isolated multitrack recordings.}

Conceptually, our approach is similar of drum replacement in music production, where drum tracks undergo manual onset detection and are replaced with one-shot samples from a sample library. We propose instead to perform this process in an automated, data-driven manner while learning to synthesize the actual one-shot samples from the mixture.

Our method, \ac{IDM}, combines approaches from \ac{ADT}, unsupervised source separation, and drum synthesis literature \cite{choiDeepUnsupervisedDrum2019, vandeveireSigmoidalNMFDConvolutional2021, shierDifferentiableModellingPercussive2023, schulze-forsterUnsupervisedMusicSource2023} to tackle \ac{DSS} without requiring separated stems, \new{only requiring transcription annotations at training time}. \new{An open-source implementation of our method is available online\footnote{\url{https://github.com/bernardo-torres/inverse-drum-machine}}}. Our main contributions are:
\new{
\begin{itemize}
\item A novel analysis-by-synthesis framework for \acf{DSS} that works without isolated stems, relying only on transcription data for training.
\item A jointly trained model that unifies \acf{ADT} and  \ac{OSS} in a single end-to-end system.
\item A modular separation model that achieves separation quality comparable to supervised, state-of-the-art methods while using $\approx$ $100$ times fewer parameters.
\end{itemize}
}

\new{The remainder of this paper is structured as follows: Section \ref{sec:related_work} reviews related work in \ac{ADT}, \ac{OSS}, \ac{DSS}, and analysis-by-synthesis. The proposed Inverse Drum Machine (IDM) is explained in Section \ref{sec:method}, and its experimental evaluation is detailed in Section \ref{sec:experiments}. We present and discuss the results in Sections \ref{sec:results} and \ref{sec:discussion} and conclude in Section \ref{sec:conclusion}.}

\section{Related Work}
\label{sec:related_work}

This section presents relevant prior research in four interconnected areas that form the foundation of our approach: \acf{ADT}, \acf{OSS} , \acf{DSS}, and Data-driven analysis-by-synthesis.

\subsection{\acf{ADT}}

\ac{ADT} aims to detect and classify percussion events within audio recordings, producing a symbolic representation (typically onset times and instrument classes) of the drum performance \cite{wuReviewAutomaticDrum2018}. While \ac{ADT} encompasses various formulations, this work focuses on \ac{DTD}, where the input consists exclusively of percussion instruments without melodic content.

Approaches to \ac{DTD} have evolved from traditional signal processing methods to machine learning techniques.  \ac{NMF} and its variants have \sout{had lots of} \new{seen considerable} success \cite{lindsay-smithDrumkitTranscriptionConvolutive2012} modeling spectral patterns of drum sounds without requiring labeled training data. More recent methods leverage deep learning, with both supervised \cite{voglDrumTranscriptionJoint2017, callenderImprovingPerceptualQuality2020} and unsupervised \cite{choiDeepUnsupervisedDrum2019} approaches showing significant performance improvements.
Most \ac{DTD} research has focused on a limited set of drum instruments, typically kick drum, snare drum, and hi-hat \cite{dittmarRealtimeTranscriptionSeparation2014, lindsay-smithDrumkitTranscriptionConvolutive2012}, while some recent works have expanded the instrument vocabulary \cite{callenderImprovingPerceptualQuality2020, choiDeepUnsupervisedDrum2019, voglMultiinstrumentDrumTranscription2018, zehrenHighqualityReproducibleAutomatic2023,weberRealtimeAutomaticDrum2024}.

\subsection{One-shot Drum Sample Synthesis}

Percussion instruments typically produce sounds with broad spectral characteristics and exponentially decaying modes \cite{cookPhysicallyInformedSonic1996}. Drum sound synthesis has a rich history in commercial music production and academic research. Traditional approaches include physical modeling \cite{cookPhysicallyInformedSonic1996,hanWAV2SHAPEHearingShape2020}, which simulates the acoustic properties of percussion instruments; spectral modeling \cite{iiiPARSHLAnalysisSynthesis1987,vermaExtendingSpectralModeling2000a}, which represents sounds as sinusoids, filtered noise, and a transient component; and sample-based synthesis, which triggers pre-recorded audio samples. 

Recent advances in generative deep learning have tackled drum sound synthesis research from a data-driven perspective, using convolutional neural networks \cite{ramiresNeuralPercussiveSynthesis2020} and generative adversarial networks \cite{aouameurNeuralDrumMachine2019, nistalDRUMGANSynthesisDrum2020, drysdaleAdversarialSynthesisDrum2020}. Other techniques employ diffusion models \cite{rouardCRASHRawAudio2021} and \ac{DDSP} \cite{shierDifferentiableModellingPercussive2023, simionatoSinesTransientNoise2025}.
With particular relevance to this work, Shier et al. \cite{shierDifferentiableModellingPercussive2023} proposed pairing a \ac{DDSP} sinusoidal plus noise synthesizer with a transient enhancing \ac{TCN} to generate drum sounds. \new{\acp{TCN} apply dilated convolutions to sequence data. This combination allows them to capture long-term dependencies with a large receptive field, making them effective for audio synthesis.}

\subsection{\acf{DSS}}

Traditional \ac{DSS} approaches have primarily relied on \ac{NMF}-based methods, which model drum events as spectral templates repeatedly triggered over time \cite{gilletTranscriptionSeparationDrum2008}. These approaches decompose a spectrogram into a product of template and activation matrices, representing the spectral characteristics of different drum sounds and their temporal occurrences, respectively. 

Conceptually, our approach shares several similarities with works based on \ac{NMFD} \cite{smaragdisNonnegativeMatrixFactor2004} which incorporate transcription information into the initialization process\cite{dittmarRealtimeTranscriptionSeparation2014, dittmarReverseEngineeringAmen2016}. However, our method differs in several crucial aspects: (1) we operate directly in the time domain, providing immediate access to one-shot samples without requiring magnitude spectrogram inversion; (2) unlike \ac{NMFD}, which \sout{updates spectral templates at inference time through multiplicative update rules} \new{performs per-track optimization on test-time}, our approach incorporates a dedicated training stage that optimizes neural network parameters through gradient descent; \new{(3) our method can bypass Wiener filtering (which typically uses the phase from the mixture) to obtain separated sources}.

In \ac{DSS} literature, \ac{NMFD} methods have also been used in a cascaded manner \cite{dittmarReverseEngineeringAmen2016} to reduce cross-talk between instruments, and tested for dual-channel input \cite{caiDualchannelDrumSeparation2021}. Vande Veire et al. \cite{vandeveireSigmoidalNMFDConvolutional2021} introduced Sigmoidal \ac{NMFD}, which enforces impulse-like behavior in the updated activations by factoring them into an onset and an amplitude component, and updating them using gradient descent.

Despite their effectiveness in controlled settings, \ac{NMFD}-based methods have limited flexibility \sout{and cannot learn from data} since they operate at test time \new{and are typically trained on a per-track basis}. Moreover, while these approaches have demonstrated reasonable performance on datasets with few instrument classes, they have not been extensively evaluated on large-scale datasets with numerous percussion types.

A significant advancement in \ac{DSS} came with the development of the StemGMD dataset \cite{mezzaDeepDrumSource2024}, which leveraged MIDI transcriptions from the GMD dataset \cite{gillickLearningGrooveInverse2019}, captured from performances on an electronic drum kit, to synthesize individual drum stems for nine drum classes. With it, \new{the first deep neural network specifically designed for drum source separation was introduced, capable of separating} five stems from a stereo drum mixture: kick drum, snare drum, tom-toms, hi-hat, and cymbals. A recent benchmark of deep \ac{MSS} architectures trained on StemGMD \cite{mezzaBenchmarkingMusicDemixing2024} shows that there is room for improvement by using larger, more complex models,  in particular those that are trained to perform waveform synthesis directly (rather than relying on time-frequency masking). Unlike \ac{NMFD}, these \ac{DNN} models require supervised training with ground-truth separated stems and often have a high parameter count, making them computationally expensive and data-intensive.

\subsection{Data-driven Analysis by Synthesis}
A growing body of research combines data-driven methods with analysis-by-synthesis approaches. \acf{DDSP} combines \new{gradient descent (particularly in automatic differentiation frameworks)} with \ac{DSP} \cite{engelDDSPDifferentiableDigital2020}, enabling end-to-end training of models that incorporate signal processing operations within their architectures. \new{This paradigm has been applied in several works for the} estimation of synthesizer parameters \cite{hanWAV2SHAPEHearingShape2020, masudaSynthesizerSoundMatching2021, shierDifferentiableModellingPercussive2023, torresUnsupervisedHarmonicParameter2024, hayesReviewDifferentiableDigital2024}.  \sout{oscillator frequency estimation, estimation of audio effects, and others}

A typical scenario is training \acp{DNN} to predict \ac{DSP} parameters for synthesis by comparing reconstructed audio to a target. Spectral loss functions, such as multi-resolution \ac{STFT} loss, are commonly used instead of waveform loss to avoid phase alignment issues.

In source separation, analysis-by-synthesis has been employed to estimate the parameters of musical sources from mixture inputs. These parameters drive source models to resynthesize individual sources that are summed and compared to the original mixture \cite{kawamuraDifferentiableDigitalSignal2022, schulze-forsterUnsupervisedMusicSource2023}. This approach has proven particularly effective for separating singing voices in choir ensembles \cite{schulze-forsterUnsupervisedMusicSource2023, richardFullyDifferentiableModel2024}, where multi-pitch information guides the separation process. 

In the context of \ac{ADT}, reconstruction of drum tracks has been used for  evaluating transcription performance through listening tests \cite{callenderImprovingPerceptualQuality2020}. For unsupervised \ac{ADT}, Choi and Cho \cite{choiDeepUnsupervisedDrum2019} resynthesize tracks from estimated transcriptions using randomly selected one-shot samples from a class-sorted collection, training a \ac{DNN} using an onset-enhanced reconstruction loss. Our work extends their method to \ac{DSS}.

\section{Method}
\label{sec:method}

\begin{figure}
    \centering
    \includegraphics[width=0.9\columnwidth,
    trim={20pt 25pt 20pt 20pt}, clip]{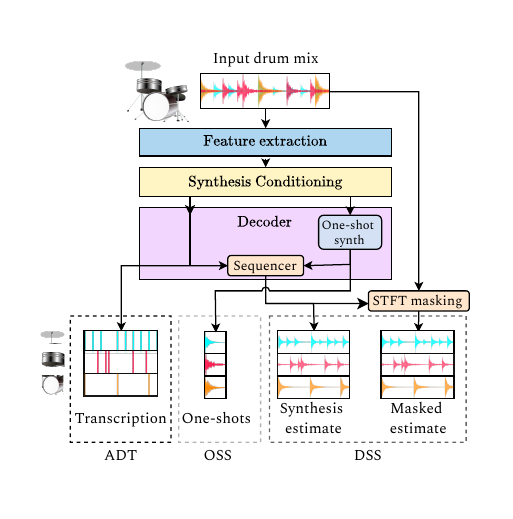}
    \caption{The proposed separation model processes the audio input through several stages. First, a Feature Extraction module extracts learned  frame-level features. These features are transformed by a Synthesis Conditioning module into relevant synthesis parameters: transcription onsets, velocities, a mixture embedding, and individual track gains. A decoder module synthesizes one-shot samples for each drum instrument conditioned on the mixture embedding and reconstructs individual drum tracks by sequencing these one-shots with the obtained transcription.  The framework encompasses three interconnected tasks: \acf{ADT}, \acf{OSS}, and \acf{DSS}. \ac{DSS} is obtained either from the decoder output  (\textit{synthesis estimate}) or after  time-frequency masking (\textit{masked estimate}).}
    \label{fig:overview}
\end{figure}

In this work, we propose an analysis-by-synthesis approach to decompose drum mixtures into transcription information and elementary one-shot samples. As illustrated in Figure~\ref{fig:overview}, the proposed architecture processes audio mixtures through several stages. Initially, a Feature Extraction module derives frame-level representations from the input mixture. These features are subsequently transformed by a Synthesis Conditioning module into relevant parameters for synthesis (transcription and one-shot conditioning). The conditioning is used by a one-shot synth to synthesize drum samples which are sequenced by the estimated transcription. \new{A multitask objective enables all modules to be trained jointly, using ground-truth transcriptions as supervision. At inference time, the model operates on the audio mixture alone; however, its modular design allows users to optionally provide external information, such as a corrected transcription.}

\begin{figure}
    \centering
    \includegraphics[width=\columnwidth,
    trim={0 0 0 0}, clip]{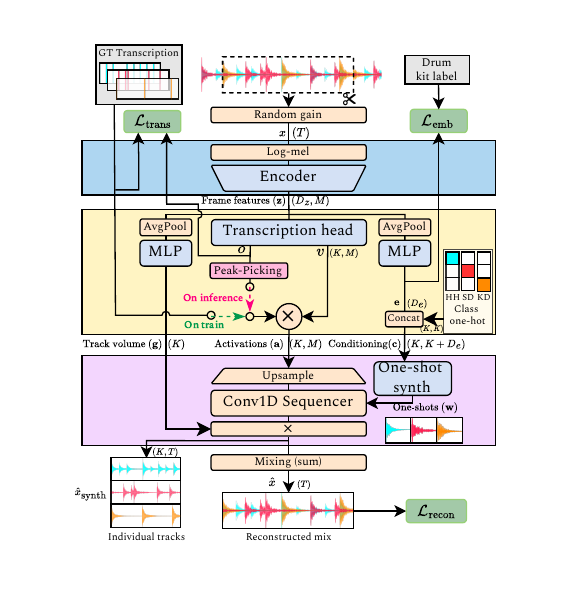}
    \caption{Detailed diagram of the analysis-by-synthesis pipeline, containing the Feature Extraction (blue), Synthesis Conditioning (yellow), and Decoder (purple) modules. Blocks in blue indicate trainable components, blocks in orange indicate differentiable operations, and blocks in pink indicate non-differentiable operations. Loss functions are represented as green components, and grey components represent external information used during training}
    \label{fig:detailed_diagram}
\end{figure}

The following sections explain the proposed multitask framework and then go into detail on the Feature Extraction,  Synthesis Conditioning, and Decoder modules, which are also detailed in Figure~\ref{fig:detailed_diagram}. For presentation clarity, we \new{leave} most implementation details to Section~\ref{sec:implementation}.

\subsection{Multitask learning for Drum Source Separation}

Our approach encompasses three interconnected tasks:

\begin{enumerate}
    \item \textbf{\acf{ADT}:} The precise estimation of the onset times of each drum instrument is achieved by training a transcription head to predict onset activations. 
    \item \textbf{\acf{OSS}:} High-quality one-shot samples for each drum instrument are generated by a \ac{TCN} conditioned on instrument type and mixture embedding.
    \item \textbf{\acf{DSS}:} Individual drum tracks are extracted from the mixture by sequencing the synthesized one-shot samples with the estimated transcription.

\end{enumerate}

\new{To train our analysis-synthesis framework we analyze an input mixture $\x$ and recompose the individual drum tracks by sequencing onset activations with generated one-shot samples. Individual tracks are mixed together to obtain a reconstructed mixture \new{$\hatxsynth$}. The entire framework is trained end-to-end with a reconstruction loss \lrecon~, with the addition of a transcription loss \ltrans~ and a mixture embedding loss \lemb~. Sections ~\ref{sec:train_iter} and~\ref{sec:training_objective} provide details on the training step and objective.}

\subsection{Feature Extraction}

Our Feature Extraction module learns features \framefeatures~which contain relevant information for the transcription and mixture embedding. It begins by computing \logmel~Spectrograms from the input waveform $\x$ at $f_s=$\samplerate. Then, a ConvNeXt encoder \cite{liuConvNet2020s2022} \new{(Section~\ref{sec:encoder_arch})} processes the spectrogram to extract frame-level representations $\framefeatures \in \mathbb{R}^{\nframefeatures \times \frames}$, where $\nframefeatures=32$ denotes the dimension of the frame features and $\frames$ represents the number of spectrogram frames.

\subsection{Synthesis Conditioning}

The Synthesis Conditioning module transforms the frame features into parameters that guide the reconstruction process: transcription onsets, velocities, individual track gains, and a conditioning vector. Estimated onsets and velocities are multiplied and upsampled to obtain an activation signal used to trigger one-shot samples, composing the individual tracks for \insts~predefined drum classes. The conditioning vector is used as conditioning for the one-shot synthesis model, guiding the generation of one-shot samples based on the instrument class and a mixture embedding, which controls the timbral variations within a given class.

\smallskip

\subsubsection{\textbf{Transcription head}} The transcription head \new{(Section~\ref{sec:transc_head_details})} transforms features \framefeatures~into a factorized representation of the drum component activation signal, composed of: (i) an onset signal $\onsets \in [0,1]^{\insts \times \frames}$, identifying when each instrument is played; and (ii) a velocity signal $\velocity \in [0,2]^{\insts \times \frames}$, capturing the intensity of each onset. This architecture draws inspiration from the "Onsets and Frames"  \cite{hawthorneOnsetsFramesDualobjective2018,callenderImprovingPerceptualQuality2020} and Sigmoidal \ac{NMFD} \cite{vandeveireSigmoidalNMFDConvolutional2021} papers. \new{We use the name velocity as a reference to \textit{MIDI velocity}, a control signal used to represent note intensity. However, we only use $\velocity$ to represent the one-shot volume and it does affect timbre.}
The temporal resolution is set by the spectrogram's hop size.

\smallskip

\subsubsection{\textbf{Activation signal}} We compose $\activation^{(\text{T})} \in \mathbb{R}^{\insts \times  \tracklength}$, the activation signal, \new{by stacking the individual instrument activations $\activation_k$, which are derived} by element-wise multiplication of onsets and velocities, followed by upsampling from the rate of the frames $m$ to the rate of the signal time $t$:
\begin{equation}
\activation^{(\text{T})}_{k} = \operatorname{upsample}(\onsets_{k} \odot \velocity_{k})
\end{equation} 

We upsample using zero insertion \cite{choiDeepUnsupervisedDrum2019}. While this approach induces aliasing, we found that if the onset signal is sparse enough, the synthesis artifacts are negligible. We therefore separate training and inference behavior to minimize these effects during training\new{.}

During {\color{OliveGreen}{training}}, we pass ground-truth onsets instead of \onsets~ to compose the activation \activation. Note that the onset signal is still used for training the transcription head from the transcription loss, but we effectively stop the gradient flowing from the reconstruction process to \onsets. The gradient passed to the velocity branch, however, remains intact. During {\color{magenta} inference}, we apply peak picking \new{(Section~\ref{sec:peak_picking})} to extract sparse onsets from  \onsets.

\smallskip



\subsubsection{\textbf{Mixture embedding}} \label{sec:mixture_embedding}
 To represent the overall timbral characteristics of the drum-kits (not of the specific instruments), we derive a mixture embedding $\embedding \in \mathbb{R}^{\nembedding}$ from frame features \framefeatures~using a \ac{MLP} (\new{please refer to Section~\ref{sec:mixture_embedding_details} for architectural details). An ideal conditioning embedding should be capable of summarizing the timbre of an instrument as well as acoustic conditions while learned without supervision. As a simpler method given the small number of drum kits used in our experiments we learn mixture embeddings by} matching one-hot labels constructed from drum kit annotations using a classification loss \lemb~\new{(Section \ref{sec:training_objective})}. This branch behaves as a "drum kit classifier" and restricts the model to synthesize only from drum kits observed during training, but serves as proof-of-concept for a disentangled class/timbre conditioning mechanism.  

\new{\subsubsection{\textbf{Track volume}} \label{sec:track_gains}
We estimate global track gains $\gain \in [0, 2]^K$ from \framefeatures~using a \ac{MLP} (Section~\ref{sec:mixture_embedding_details}).
}

 \subsubsection{\textbf{Conditioning vector}} \label{sec:conditioning_vector}
 
 Recall that  \insts~is the number of drum instruments.  A class conditioning vector \conditioning~is composed by broadcasting \embedding~to $\new{(}\insts, \nembedding\new{)}$ and concatenating ($\oplus$) the $K$-dimensional one-hot encodings of the different classes in the feature dimension, resulting in size $\new{(}\insts, \insts + \nembedding\new{)}$. \conditioning~can be interpreted as the stack of individual class\new{/}mixture conditioning vectors for all classes of a particular mixture ($\{ \conditioning_k \}_{k=1}^{\insts}$, where $\conditioning_k = \embedding \oplus \text{one-hot}(k)$).

\subsection{Decoder}

The Decoder module synthesizes one-shot samples for each drum instrument and reconstructs individual drum tracks by sequencing these one-shots with the activations from the Synthesis Conditioning module. The decoder is composed of two main components: a one-shot synth and a sequencer.

\subsubsection{\textbf{One-shot Synth}} \label{sec:one-shot}

\begin{figure}
    \centering
    \includegraphics[width=\columnwidth,
    trim={20pt 12pt 20pt 10pt}, clip]{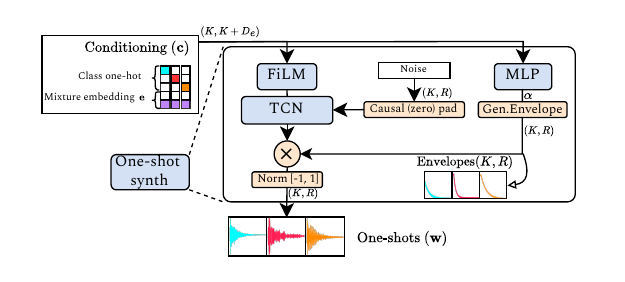}
    \caption{One-shot synth model architecture. White noise is fed to a \acf{TCN} conditioned via \acf{FiLM} on a conditioning vector \conditioning, which has disentangled instrument class/timbre dimensions. \new{Causal zero padding is applied}. \new{The output is shaped by a parametrized exponential envelope estimated from $\conditioning$ and normalized to $[-1, 1]$ amplitude range.  $\conditioning$ controls the drum kit (mixture embedding) and drum instrument (class one-hot) synthesized by the \ac{TCN}}.}  
    \label{fig:one_shot_synth}
\end{figure}

Conditional \acl{OSS} is performed with a \acf{TCN}, similar to the work of Shier et al. \cite{shierDifferentiableModellingPercussive2023}. However, instead of using \ac{TCN} to enhance the transient of a sinusoidal-plus-noise signal \cite{vermaExtendingSpectralModeling2000a}, we take a more direct path.
We eliminate the sinusoidal-plus-noise input, instead feeding white noise directly to the \ac{TCN}. By expanding the network's receptive field and adding a time envelope, we found the \ac{TCN} capable of synthesizing both sinusoidal, noise, and transient components. \new{Importantly, in order to assure the TCN output has length of \sampleduration~second, the input white noise is left zero-padded accordingly. This causal padding is also essential for modeling the transient. By providing a block of zeros before the noise, we encourage the TCN to not operate in a continuous steady-state mode during the transient period.} Figure~\ref{fig:one_shot_synth} shows the architecture of the one-shot synthesis model, with specific implementation details reported in Sections~\ref{sec:tcn_architecture} to~\ref{sec:envelope}.

To guide the synthesis process, we \new{follow \cite{shierDifferentiableModellingPercussive2023}} and condition the \ac{TCN} using  \ac{FiLM} \cite{perezFiLMVisualReasoning2017} on a conditioning vector \conditioning~composed of two disentangled dimensions: a mixture embedding \embedding~and the instrument class one-hot encoding. To model the natural energy decay typical of percussion instruments, an exponential amplitude envelope is multiplied to the \ac{TCN} output, controlled by a \new{decay} parameter $\alpha$ dependent on \conditioning. 

Unlike most \ac{OSS} approaches \cite{nistalDRUMGANSynthesisDrum2020,shierDifferentiableModellingPercussive2023,rouardCRASHRawAudio2021}, we train the synthesizer  model on mixture \new{track} reconstructions rather than isolated one-shot samples, making it inherently reliant on accurate onset placement.

\smallskip

\subsubsection{\textbf{Sequencer}}
    The sequencer serves as the bridge between transcription and synthesis, transforming drum instrument activations and one-shot samples into full individual tracks. Let $\samples \in \new{[-1,1]}^{\insts \times \onesecondsamples  }$ represent a set of time-domain one-shot signals (one second in duration \sout{, \new{$\onesecondsamples=f_s$}),  $\gain \in \mathbb{R}^K$ denote a vector of global track gains}. The reconstructed mixture $\hatxsynth \in \mathbb{R}^{T}$ is obtained by:

\begin{equation}
    \hatxsynth = \sum_{k=1}^{\insts} \gain_k \sequencer(\samples_k, \activation^{(\text{T})}_k)
\end{equation} \label{eq:reconstruction}

The sequencer is implemented as a convolutional operator using frequency domain multiplication \cite{choiDeepUnsupervisedDrum2019}.

\subsection{Time-frequency masking}

Although separated sources can be directly obtained by synthesis, previous work showed that improved results can be obtained after using them for spectral masking of the mixture \cite{schulze-forsterUnsupervisedMusicSource2023}. In this work, we employ $\alpha$-Wiener Masking \cite{liutkusGeneralizedWienerFiltering2015} \new{(during inference only)}, which uses our synthesized individual tracks to create \ac{TF} masks applied to the original mixture. The underlying principle is that each stem's spectrogram contributes proportionally to the energy in each \ac{TF} bin of the mixture's spectrogram (\new{see Section \ref{sec:masking} for details)}.

\subsection{Implementation details} \label{sec:implementation}

\subsubsection{\textbf{Log Mel Spectrogram}} We compute Log Mel Spectrograms with \new{$128$} mel frequency bands, a window size of \new{$2048$} samples, and a hop size of \new{$512$} samples (\new{\hopsizems ms)}, corresponding to \new{\activationrate} Hz activation rate. 
\smallskip

\subsubsection{\textbf{Encoder Architecture}}\label{sec:encoder_arch} Our ConvNeXt encoder begins with a transition convolutional layer downsampling the feature dimension by 4, followed by \texttt{LayerNorm} \cite{baLayerNormalization2016}. Four hierarchical stages follow with feature dimensions $[4,8,16,\nframefeatures]$, each starting with a channel downsampling block (\texttt{LayerNorm} + $3\times3$ convolution). Each stage contains multiple residual blocks with $7\times7$ depthwise separable convolutions, \texttt{LayerNorm}, \texttt{GELU} activations, and pointwise transformations. Unlike the original ConvNeXt, we maintain temporal resolution between stages. As demonstrated in \cite{zehrenHighqualityReproducibleAutomatic2023}, the specifics of the architecture might not have a big impact on \ac{ADT} performance. We thus opted for a modern, well-validated off-the-shelf convolutional architecture.

\smallskip

\subsubsection{\textbf{Transcription Head}}\label{sec:transc_head_details}

The frame features \framefeatures~are processed through a 1D convolutional layer with \texttt{ReLU} activation, followed by parallel onset and velocity branches with linear layers. The onset branch outputs sigmoid activations, while the velocity branch uses exponentiated sigmoid activation \cite{engelDDSPDifferentiableDigital2020}. We did not include a sequence model as we found it unnecessary for achieving reasonable transcription performance.

\smallskip

\subsubsection{\textbf{Peak Picking}} \label{sec:peak_picking}
 We apply peak picking to the frame-rate onset signal during inference to extract sparse onsets and remove low amplitude activations. We follow the heuristic \new{and hyperparameters} from \cite{bockEvaluatingOnlineCapabilities2012,choiDeepUnsupervisedDrum2019}. 
\smallskip

\subsubsection{\textbf{Mixture Embedding \new{and Track gain MLPs}}} \label{sec:mixture_embedding_details}

\new{Both \ac{MLP} architectures first remove the time dimension from the frame features \framefeatures~via temporal average pooling. The Mixture Embedding \ac{MLP} then applies a \texttt{LayerNorm} layer, followed by a two-layer \texttt{ReLU} network that maps dimensions from $D_z \rightarrow D_z \rightarrow \nembedding~$. For the track gains, a separate three-layer \ac{MLP} with \texttt{LeakyReLU} activations is applied per instrument, with layer dimensions mapping from $64 \rightarrow 16 \rightarrow 16 \rightarrow 1$. The final layer uses an exponentiated sigmoid activation~\cite{engelDDSPDifferentiableDigital2020} to estimate each gain parameter~$\gain_k$.}

\smallskip

\subsubsection{\textbf{TCN Architecture}}\label{sec:tcn_architecture} Our TCN \cite{shierDifferentiableModellingPercussive2023} consists of \numlayers~dilated causal convolutional layers with residual connections and \texttt{GELU} activations. Input is projected to a latent dimension of $48$, processed through the dilated residual blocks (dilation doubling at each layer, starting from $1$), and projected back to audio dimensions. The total receptive field is \new{$\approx1.3$s}. Kernel size is \kernelsize~throughout, with instance normalization applied to all layers and output normalized to $[-1,1]$.

\smallskip

\subsubsection{\textbf{FiLM Conditioning}}\label{sec:film} Conditioning is implemented as a stack of linear layers mapping from \conditioning~to shift/scale parameters applied after the \texttt{GELU} activation in each \ac{TCN} layer:

\begin{equation}
    \text{FiLM}(\mathbf{x}_{ij}, \conditioning) = \gamma_{ij}(\conditioning) \cdot \mathbf{x}_{ij} + \beta_{ij}(\conditioning) ,
    \end{equation}

\noindent where $\mathbf{x}_{ij}$ is the $j\text{th}$ feature map of the \new{$i\text{th}$}  \ac{TCN} layer (after \texttt{GELU}), $\conditioning$ is the conditioning vector, and $\gamma_{ij}(c)$ and $\beta_{ij}(c)$ are the shift and scale parameters learned by the $(i,j)\text{th}$ linear layer.

\smallskip

\subsubsection{\textbf{Envelope}}\label{sec:envelope} A $2$ layer \ac{MLP}  with \texttt{ReLU} activation and exponentiated sigmoid estimates envelope parameter $\alpha$ from the conditioning vector. The envelope is applied to the \ac{TCN} output by element-wise multiplication:
\begin{equation}
    \text{Gen. envelope}(\alpha) = e^{- 20 \cdot \alpha \cdot \frac{t}{\onesecondsamples}} \quad \text{for} \new{\quad t \in \mathbb{Z}, 0 \le t < \onesecondsamples}
\end{equation}

\subsubsection{\textbf{TF Masking}}\label{sec:masking} Given a mixture complex \ac{STFT} $\mathbf{X}$ and an estimated source  \ac{STFT} $\mathbf{\hat{X}}_{\text{synth},i}$, time-frequency masks $\mathbf{M}_i$ and masked spectrograms $\mathbf{\hat{X}}_{\text{mask},i}$ are computed as:

\begin{equation}
    \new{\mathbf{M}_i = |\mathbf{\hat{X}}_i|^\alpha \oslash (\sum_j |\mathbf{\hat{X}}_j|^\alpha + \epsilon) }\quad \text{and} \quad
    \mathbf{\hat{X}}_{\text{mask},i} = \mathbf{M}_i \odot \mathbf{X},
\end{equation}
\noindent \new{where $\odot$ and $\oslash$ denote element-wise multiplication and division, respectively}, and $\epsilon$ is a small constant to prevent division by zero. We denote $\hat x_\text{mask}$ the waveform estimates obtained after inverse \ac{STFT} of $\hatX_{\text{mask},i}$ using the mixture phase. $\alpha$ is set to $1$ in order to remain consistent with  \cite{dittmarReverseEngineeringAmen2016,mezzaDeepDrumSource2024}.


\section{Experiments}\label{sec:experiments}

This section describes our experimental methodology for training and testing the proposed drum source separation framework. 
Our primary experimental configuration uses a drums-only database with ground-truth transcription annotations and separated stems as a test-set. The following subsections detail the data, experimental protocols, \new{hyperparameters}, and baseline methods employed in our work.

\subsection{Datasets} 

Our experiments are conducted using the StemGMD dataset \cite{mezzaDeepDrumSource2024}, which builds upon the MIDI data from the GMD dataset \cite{gillickLearningGrooveInverse2019} by synthesizing isolated drum stems for $9$ canonical instruments across $10$ distinct drum kits. StemGMD has approximately $136$ hours of mixture data.  Perfect transcription annotations were extracted directly from the associated MIDI files. Table~\ref{tab:class_mapping} presents the mapping between the original instrument nomenclature, the names used in the audio files (\texttt{9-Class}), and the two instrument groupings employed in our experimental design. Training uses the \texttt{9-Class} configuration, while evaluation is performed on both \texttt{9-Class} and \texttt{5-Class} groupings. To maintain comparability with  \new{baseline} \larsnet~\new{(Section \ref{sec:baselines})}, which withheld $4$ kits during training, we use only $6$ of the $10$ drum kits available in the dataset (\new{$\nembedding=6$}).

\begin{table}
    \centering
    \caption{Drum class configurations. Instrument names correspond to common drum kit terminology, while \texttt{9-Class} and \texttt{5-Class} refer to the groupings used in our experiments. \texttt{9-Class} naming is consistent with the StemGMD dataset. Our model is trained on the \texttt{9-Class} configuration and is able to separate an input mixture into $9$ drum classes. The \texttt{5-Class} configuration, used for evaluation only, groups similar instruments together.}
    \begin{tabular}{lll}
    \toprule
    Instrument Name & \texttt{9-Class} (train/test) & \texttt{5-Class} (test) \\
    \midrule
    Kick Drum & kick & Kick (KD)\\
    Snare Drum & snare & Snare (SD) \\
    \midrule
    Closed Hi-Hat & hihat\_closed & \multirow{2}{*}{Hi-Hats (HH)} \\
    Open Hi-Hat & hihat\_open & \\
    \midrule
    High-Mid Tom & hi\_tom & \multirow{3}{*}{Toms (TT)} \\
    Low Tom & mid\_tom & \\
    High Tom & low\_tom & \\
    \midrule
    Crash & crash\_left & \multirow{2}{*}{Cymbals (CY)} \\
    Ride & ride & \\
    \bottomrule
    \end{tabular}
    \label{tab:class_mapping}
\end{table}

We follow the provided train, validation, and test splits on the MIDI track level. Contrary to \larsnet \cite{mezzaDeepDrumSource2024}, which uses a small subset of the test partition named \textit{eval session}, we evaluate our models on the entire test split, which comprises $\sim~10$ h of mixture data. We found that  the \textit{eval session} removed substantially the toms and cymbals from the test set, and by increasing its size we achieve a more balanced evaluation. 

\new{\subsection{Training}\label{sec:train_iter}}

Our training protocol uses randomly cropped audio segments of $T=\new{\duration}$~seconds \new{at sampling rate $f_s=$ \samplerate}, normalized to the amplitude range $[-1, 1]$.   Cropped tracks are resampled and converted to mono on-the-fly. We apply random gain augmentation to \new{obtain input $\x$}, with gain values sampled from a uniform distribution in the range $[0.3, 1.0]$ with a probability of $0.8$.  \new{We use no additional data augmentations.}

\new{Transcription activations and a conditioning vector (\conditioning) are extracted from $\x$. Then,  individual one-shot samples (\oneshots) for each drum instrument are synthesized from \conditioning~and sequenced into full tracks using the activations. The individual tracks are mixed to obtain a reconstructed mixture $\hat{x}$, and reconstruction loss \lrecon~ is applied between $\hat{x}$ and $\x$. The entire framework is trained end-to-end, with the addition of a transcription loss \ltrans~ and a mixture embedding loss \lemb~ (Section~\ref{sec:training_objective}). }

For the training phase, we configure the number of drum instruments as $K=9$, enabling separation of the input mixtures into $9$ distinct drum classes (as enumerated in Table~\ref{tab:class_mapping}). Training consists of $800$ epochs with a batch size of $33$, using the Adam optimizer with a learning rate of $5 \times 10^{-3}$ and gradient clipping with a threshold value of $0.5$. We implement model checkpointing based on the mean validation loss $\mathcal{L}_{\text{total}}$ across tracks (no cropping is performed on the validation data). We keep only the shortest $550$ validation tracks for a more efficient validation. \new{Our total parameter count is of \nparams.}

\subsection{Training Objective} \label{sec:training_objective}

The system is trained end-to-end using a combination of reconstruction, transcription, and mix embedding losses. The reconstruction loss $\lrecon$ is computed between the synthesized mixture and the input cropped mixture. We set  $\lrecon$ to be the Multi-Resolution \ac{STFT} Loss \cite{engelDDSPDifferentiableDigital2020, wangNeuralSourcefilterWaveform2019}:

\begingroup
\begin{multline}
    \lrecon(\x,  \new{\hatxsynth}) = \sum_{\gamma \in \Gamma} \left\||\mathbf{X}^{(\gamma)}| - |\mathbf{\hat{X}}^{(\gamma)}|\right\|_{1} + \\ \left\|\log (|\mathbf{X}^{(\gamma)}|) - \log (|\mathbf{\hat{X}}^{(\gamma)}|)\right\|_{1},
\end{multline}
\endgroup

\noindent where ($\x$, \new{$\hatxsynth$}) are the time domain input and the reconstructed signals and \new{$\mathbf{X}^{(\gamma)} = \text{STFT}^{(\gamma)}(x)$} is the \ac{STFT} of $\x$ at scale $\gamma$. The set of scales $\Gamma$,  expressed in number of samples of the analysis window, is set to $[2048, 1024, 512, 256]$, and the hop sizes are set to a quarter of the analysis window size.

 Transcription loss $\ltrans$ is the Binary Cross-Entropy loss between the estimated onsets and the ground-truth onsets \new{for all drum instruments}. \new{Since ground-truth annotations are available as a list of onset times, in order to obtain onset target signals for training, we create a sequence containing $\frames$ zeros (in the same rate as the activation rate) and set frames closest to the annotation time to $1$}.  The mixture embedding loss $\lemb$ \new{is essentially a drum kit classification loss, implemented as the} Cross-Entropy between the estimated mixture embedding \embedding~and the one-hot drum kit label.  The training objective is the sum of the three losses:

 \begin{equation}
     \LL{total} = \lrecon + \ltrans + \lemb
 \end{equation}
 
\subsection{Baselines}\label{sec:baselines}

\begin{table}
    \centering
    \caption{Comparison of different methods for drum source separation, baselines and proposed model. Transc. = Transcription.}
    \setlength{\tabcolsep}{4pt}
    \begin{tabular}{lp{1.8cm}p{2.4cm}c}
    \toprule
    & \multicolumn{2}{c}{External information used} & \multirow{2}{*}{\parbox{1.7cm}{Input for eval.\\(kHz/channels)}} \\
    \cmidrule(lr){2-3}
    Method & Training & Inference & \\
    \midrule
    \oracle  & -- &  Isolated stems & 44.1/mono \\
    \midrule
    \nmfdonea & -- & Transc. $+$ one-shots & 44.1/mono \\
    \nmfdoneb & -- & Transc. $+$ one-shots & 44.1/mono \\
    \nmfdthree & -- & Transc. & 44.1/mono \\
    \midrule
    \larsnet & Isolated stems & -- & 44.1/stereo \\
    \larsnetmono & Isolated stems & -- & 44.1/mono \\
    \midrule
    \ours \phantom{ } (ours) & Transc. & -- & 44.1/mono \\
    \oursonsets & Transc. & Transc. & 44.1/mono \\
    \bottomrule
    \end{tabular}
    \label{tab:model_comparison}
\end{table}

We evaluate our approach against both learning-free\footnote{\new{We consider a method learning-free if it does not have a data-driven training stage designed to generalize to unseen data at test-time.}} baselines and a deep supervised baseline. Table~\ref{tab:model_comparison} summarizes our experimental comparison and the varying levels of external information used during training and inference phases.
\smallskip

\subsubsection{\textbf{Learning-free}}
The learning-free baseline methodology was proposed in 
\cite{dittmarRealtimeTranscriptionSeparation2014, dittmarReverseEngineeringAmen2016}. This approach is based on \ac{NMFD} \cite{smaragdisNonnegativeMatrixFactor2004}, wherein the mixture's spectral magnitude is modeled as the product of template and activation matrices. Their work employs informed \ac{NMFD} variants where drum transcription information and spectral templates from drum one-shots are used for \ac{NMFD} algorithm initialization. 
We replicate experimental cases $1$A, $1$B, and $3$ from their work for direct comparison with our method. 
In all three cases, the activation matrix is initialized with  ground-truth transcription information
 with values of $1$ where onsets occur and a small constant $\epsilon=10^{-10}$ elsewhere. 

Following  \cite{mezzaDeepDrumSource2024}, we use a StemGMD partition containing isolated one-shots synthesized with identical drum kits for template initialization for $1$A and $1$B (\textit{single hits} partition). In case $3$, spectral templates are initialized with random values. Both $1$A and $1$B use transcription and template initialization, but in $1$A, spectral templates remain fixed while only the activation matrix is updated. We denote these baselines by \nmfdonea, \nmfdoneb, and \nmfdthree, respectively.

We adapted the \ac{STFT} parameters to accommodate our \samplerate~ sample rate while maintaining time resolution and window size consistency with the original implementation. In contrast to findings reported in \cite{mezzaDeepDrumSource2024}, we observed a drop in performance when increasing the number of iterations. Consequently, we conducted a grid search on the validation set, varying template length and iteration count for each case. Our hyperparameter search determined that $50$ iterations resulted in optimal performance for the case with fixed templates (\nmfdonea), while $20$ iterations (\new{\texttt{NMFD} updates}) were best for \nmfdoneb~ and \nmfdthree. For the template length, we found that $40$, $10$, and $7$ yielded the best results respectively. We increased $\epsilon$ to $10^{-10}$ to address convergence issues observed with smaller values. The number of templates was configured to match the number of drum instruments ($9$). All other implementation details were preserved as specified in the original publication. \texttt{NMFD} baselines were reimplemented using TorchNMF\footnote{\url{https://github.com/yoyolicoris/pytorch-NMF}} \new{following an open source reference code \cite{lopez-serranoNMFToolboxMusic2019}}.

\smallskip
\subsubsection{\textbf{Supervised DNN}} 
We benchmark our method against the deep learning model \larsnet, which was trained for drum separation in a fully supervised manner using isolated stems \cite{mezzaDeepDrumSource2024} and its available open source\footnote{\url{https://github.com/polimi-ispl/larsnet}}. \larsnet~was trained on StemGMD multitrack data, using $6$ of the $10$ available drum kits. The model has $49.1$M parameters, operates at $44.1$ kHz, and was trained on stereo files to separate stems in the \texttt{5-Class} configuration. To ensure a fair comparison, we provide the model with \new{$44.1$ kHz stereo inputs to obtain stereo separated estimates},
\sout{Then, to compute the performance metrics, we downsample the separated stems to $16$ kHz} \new{then} convert them to mono. When evaluated in the \textit{masking} configuration, we use these estimates to compute $\alpha$-Wiener masks and apply them to the mixture spectrogram. We also report metrics when removing spatial information from the input (\larsnetmono) \new{by averaging the stereo channels and duplicating the mono input.}

\smallskip

\subsubsection{\textbf{Onsets on inference time}}
 Due to our modular synthesis approach, we can override the class onset predictions $o_k[m]$ during inference with the ground-truth (while keeping velocities  $v_k[m]$). This provides an upper performance bound by isolating transcription limitations from synthesis capabilities. This approach offers insights into the separation performance with perfect onset detection. We add \texttt{+ onsets} to the model name when reporting these results.

\subsection{Evaluation} \label{sec:evaluation}

We implement a comprehensive evaluation framework combining transcription and separation metrics. It is noteworthy that the \ac{DSS} task encompasses multiple drum classes with predominantly sparse activations (such as tom-toms), requiring careful consideration in the evaluation protocol. All metrics are computed per class, using full tracks as input, and aggregated across tracks. For all metrics except \ac{PES}, evaluation is restricted to active stems, defined as stems containing at least one onset during the track duration. This methodological decision results in variable sample sizes (number of tracks) for the different classes. The "Overall" metrics reported are computed as the aggregation of scores across all tracks, flattened and independent of class. We found this methodology to best represent the actual class distribution in the dataset.

During evaluation, input mixtures are generated dynamically through summing constituent stems. For the \texttt{5-Class} evaluation configuration, we sum the stems within the same instrument groups to establish the target stems (e.g., \texttt{hihat\_closed} and \texttt{hihat\_open} are combined into a single \texttt{Hi-Hats} stem). We also sum model outputs accordingly.

\smallskip

\subsubsection{\textbf{Transcription}}

We extract discrete onsets from the estimated onset signal through peak picking, following the methodology detailed in Section~\ref{sec:peak_picking}. Precision and recall metrics are then computed for each drum class using the \textit{mir\_eval} python package \cite{raffelMIR_EVALTransparentImplementation2014}.
The precision metric quantifies the proportion of detected onsets that correspond to actual events, while recall measures the proportion of actual events that are successfully detected by the system. 

Transcription evaluation is performed exclusively on our model \sout{and one \texttt{NMFD} baseline}. We emphasize that our transcription performance assessment is not intended as a direct comparison with state-of-the-art drum transcription systems (which typically incorporate multiple datasets and extensive augmentation strategies), but rather as a measure of our system's intrinsic transcription capabilities and associated limitations. Moreover, even though \ac{DTD} is considered a relatively simple task in \ac{ADT} literature \cite{wuReviewAutomaticDrum2018,zehrenHighqualityReproducibleAutomatic2023}, most works do not consider the full range of drum instruments present in our dataset.

\smallskip

\subsubsection{\textbf{Source separation}}

Our separation evaluation methodology is designed to assess both direct synthesis and masking-based separation capabilities. For each input mixture $\x$, the model generates synthesized estimates $\hatxsynth$ directly produced by our generative model. 

We also compute masked estimates \hatxmask~through time-frequency masking of the input mixture (Section~\ref{sec:masking}). While masking techniques have been extensively employed in source separation literature, they exhibit known limitations including spectral leakage artifacts. Synthesis, conversely, provides a more direct assessment of the model's capacity to generate authentic drum sounds but may suffer from phase misalignment with the ground-truth. We therefore evaluate synthesis and masking outputs independently.

For each active stem, we compute both waveform and perceptual metrics between model outputs and ground-truth stems, with waveform metrics being the standard in source separation literature \cite{defossezHybridSpectrogramWaveform2021, mitsufujiMusicDemixingChallenge2021}.  

\new{
To quantify separation quality in the waveform domain, we use the Signal to Noise Ratio, or \acf{nSDR} as defined in\cite{defossezHybridSpectrogramWaveform2021,mezzaDeepDrumSource2024}, computed exclusively for masked outputs ($\hat{x}_\text{mask}$):

\begin{equation}
    \text{nSDR} = 10 \log_{10}\left(\frac{\|s\|^2}{\| s - \hat{s}\|^2 + \epsilon} + \epsilon \right),
\end{equation}
}


\noindent\new{where $\epsilon=10^{-8}$}. However, spectral similarity often correlates more strongly with human perceptual judgments than signal-based methods \cite{torcoliObjectiveMeasuresPerceptual2021}. We therefore employ the \acf{LSD} as proxy for perceptual similarity:


\new{
\begin{equation}
\label{eq:lsd}
\text{LSD}(s, \hat s) = \frac{1}{T} \sum_{t=1}^{T} \sqrt{\frac{1}{F} \sum_{f=1}^{F} \left( \log \frac{|\mathbf{\hat S}_{t,f}|^2 + \epsilon}{|\mathbf{S}_{t,f}|^2 + \epsilon} \right)^2}
\end{equation}
}

\noindent where $\mathbf{S}$=$\text{STFT}(s)$ with window length of $2048$ and hop length of $512$, and $s$ and $\hat s$ are the estimated and ground-truth signals.

\begin{figure}
    \centering
    \includegraphics[width=\columnwidth,
    trim={5pt 5pt 15pt 5pt}, clip]{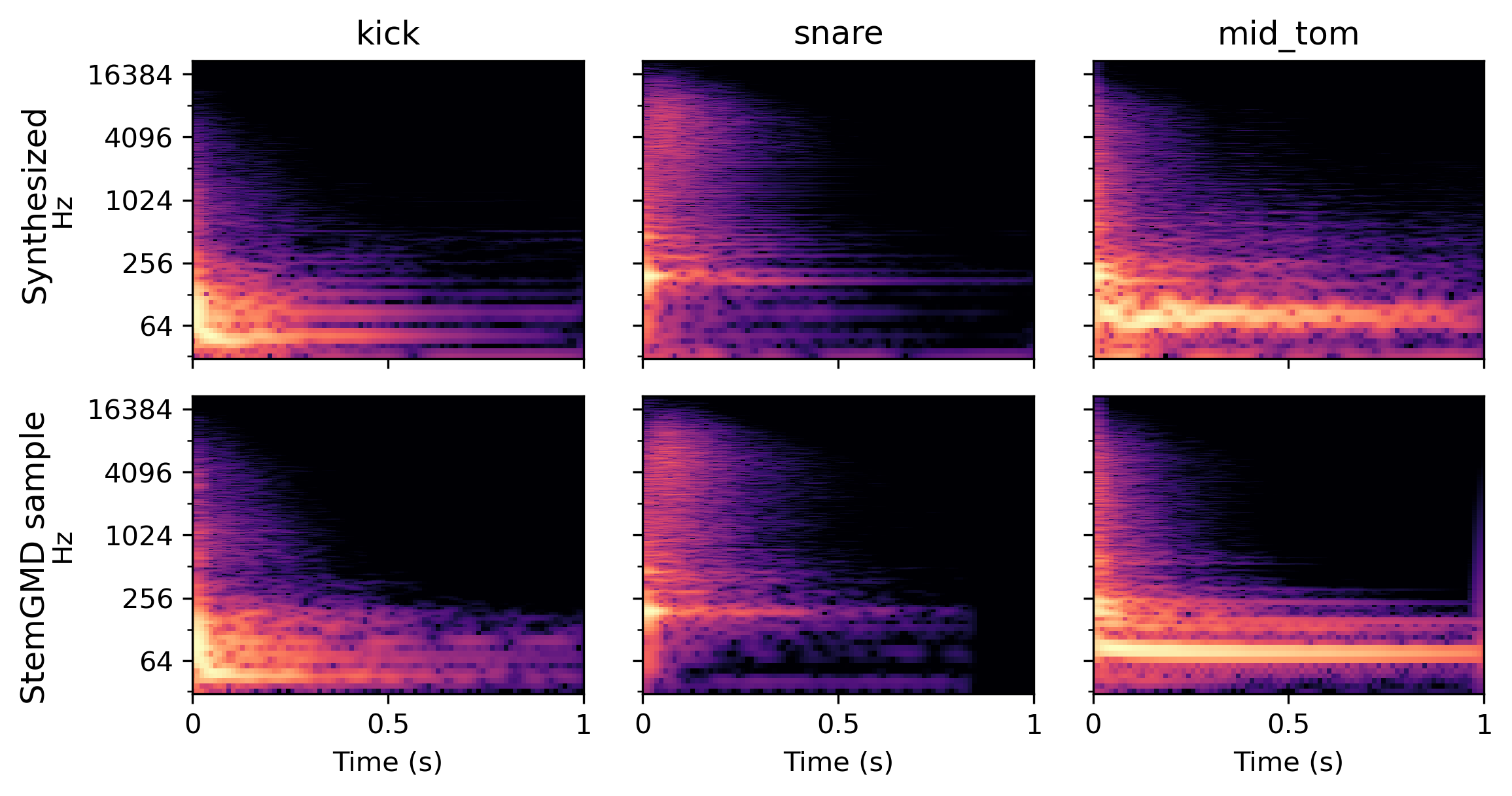}
    \caption{Log-magnitude spectrograms of synthesized, one-second-long one-shot synthesized \new{(top) and real (bottom) samples for three instruments}. The real samples are taken from the StemGMD \textit{single hits} partition with the second-highest velocity. Y axis is scaled in log frequency for better visibility, and warm\new{er} colors represent higher intensity.}
    \label{fig:side_by_side}
\end{figure}

\acf{PES} \cite{schulze-forsterWeaklyInformedAudio2019} is employed to quantify how accurately models predict silence when the corresponding ground-truth stem is silent (inactive). 
\ac{PES} is defined as the average energy of estimated stems during silent ground-truth frames. 
The estimated and ground-truth signals are segmented using non-overlapping windows of size $512$, retaining only frames corresponding to inactive ground-truth segments. 
Energy (in decibels) of estimated stems is computed as $10 \times \log_{10}(\sum_n(x(n)^2) + 10^{-8})$. We set our minimum threshold for silent frames to $-60$dB, clipping values below. 
\ac{PES} is \new{more relevant} for synthesized outputs \new{than for masked estimates}, as we observed that \ac{TF} masking introduce cross-talk artifacts that manifest as unwanted energy during silence. 

For \ac{nSDR}, higher is better, while for \ac{LSD} and \ac{PES}, lower is better.



\section{Results}
\label{sec:results}

In this section, we present a comprehensive evaluation of our proposed \acf{IDM} method and baselines, analyzing both transcription (Section~\ref{sec:transcription_results}) and separation (Section~\ref{sec:separation_results}) performance. In addition to the objective metrics shown in this paper we have made available in an accompanying website\footnote{\url{https://bernardo-torres.github.io/projects/inverse-drum-machine/}} examples of the separated stems for all models discussed, along with the synthesized one-shot samples of our model.

For box plots, the central mark indicates the median of scores across tracks, the edges of the box are the $25$th and $75$th percentiles, and the whiskers extend to the full range of the data points.

\subsection{One-Shot sample synthesis}

Our model effectively learned to synthesize drum one-shots from $9$ instruments across $6$ drum kits. We provide a qualitative analysis of a few synthesized one-shots and refer the reader to the supplementary material for a full set of samples. Figure~\ref{fig:side_by_side} shows examples of synthesized one-shots for a \new{kick}, snare and a tom-tom, with a reference one-shot taken from the StemGMD \textit{single hits} partition (second highest velocity). The generated samples demonstrate the model's ability to generate realistic drum sounds, capturing transient and steady-state components of each instrument. 

    \begin{figure}
        \centering
        \includegraphics[trim={8pt 5pt 8pt 4pt}, clip, width=\columnwidth]{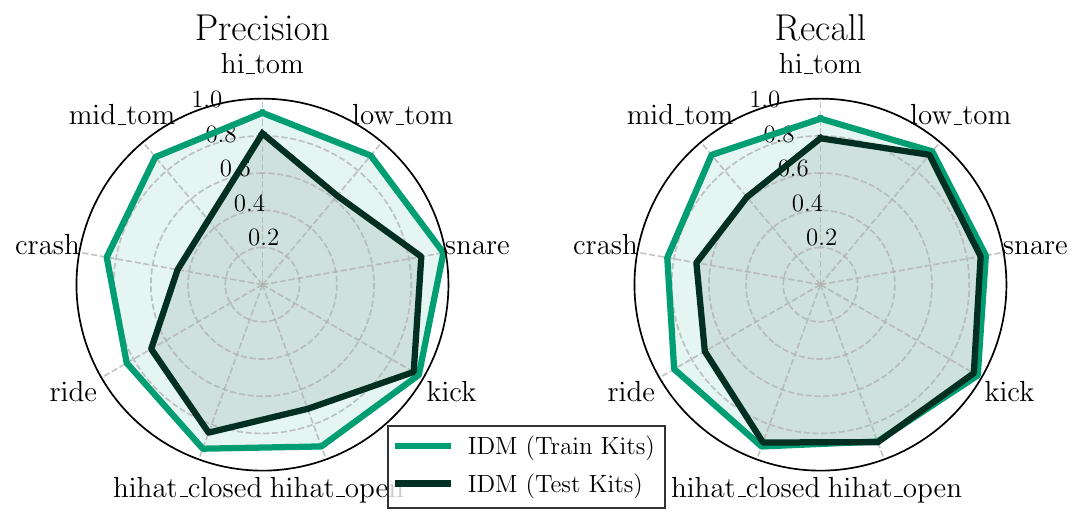}
        \caption{Performance \new{of the transcription module}.}
        \label{fig:transcription_results}
    \end{figure}
    
\begin{figure}
    \centering
    \newcommand{\HalfColWidth}{0.493\columnwidth}
    \begin{minipage}[t]{\HalfColWidth}
        \centering
        \small{\textbf{\quad \quad Train kits}}
        \subfloat{
            \includegraphics[trim={7pt 0pt 6pt 5pt}, clip, width=\textwidth]{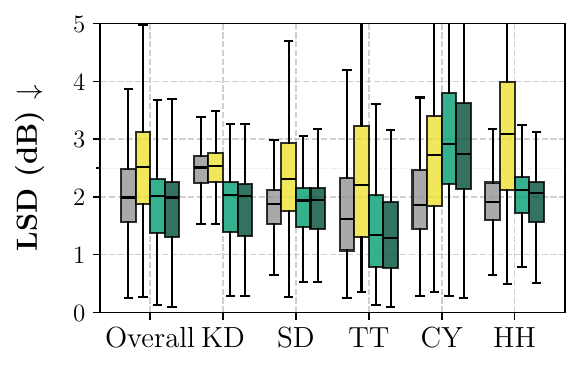}
            \label{fig:lsd_synth_train_final}
        }
        \\[-1.5mm]
        \subfloat{
            \includegraphics[trim={7pt 5pt 6pt 0pt}, clip, width=\textwidth]{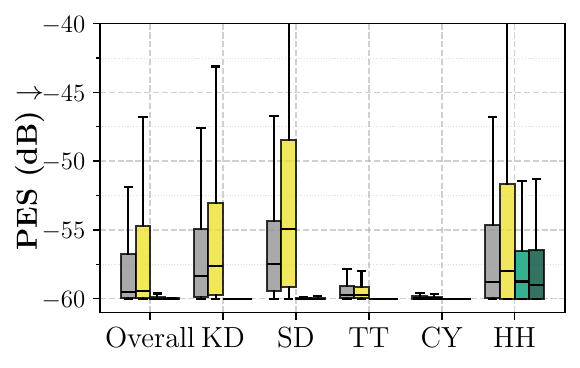}
            \label{fig:pes_synth_train_final}
        }
    \end{minipage}
    \hfill
    \begin{minipage}[t]{\HalfColWidth}
        \centering
        \small{\textbf{\quad \quad Test kits}}
        \subfloat{
            \includegraphics[trim={7pt 0pt 6pt 5pt}, clip, width=\textwidth]{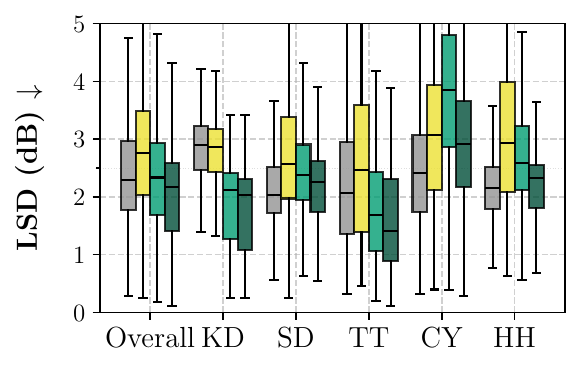}
            \label{fig:lsd_synth_test_final}
        }
        \\[-1.5mm]
        \subfloat{
            \includegraphics[trim={7pt 5pt 6pt 0pt}, clip, width=\textwidth]{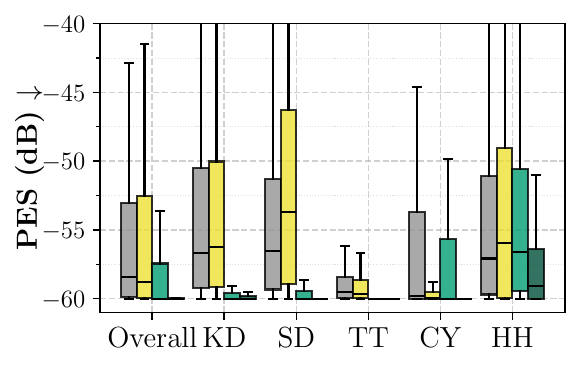}
            \label{fig:pes_synth_test_final}
        }
    \end{minipage}
    \includegraphics[width=0.75\columnwidth]{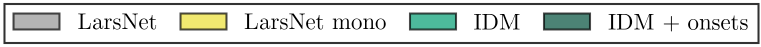}
    \caption{Comparison of synthesis-based separation metrics. Top row shows \acl{LSD}; bottom row shows \acl{PES} (lower is better for both). Left column is train kits; right is test kits.}
    \label{fig:synth_results_final}
\end{figure}

    \subsection{Transcription Results} \label{sec:transcription_results}

\newcommand{\coltitlespace}{\vspace{0mm}}
\newcommand{\LeftColImgWidth}{8}  
\newcommand{\RightColImgWidth}{7}
\newcommand{\LeftColWidth}{0.55\textwidth}
\newcommand{\RightColWidth}{0.43\textwidth}

\begin{figure*}
    \centering
    \begin{minipage}[t]{\LeftColWidth}
        \centering
        \small{\quad \quad\textbf{Train kits}}
        \subfloat{
            \includegraphics[ trim={0 4pt 0 7pt}, clip, width=\textwidth]{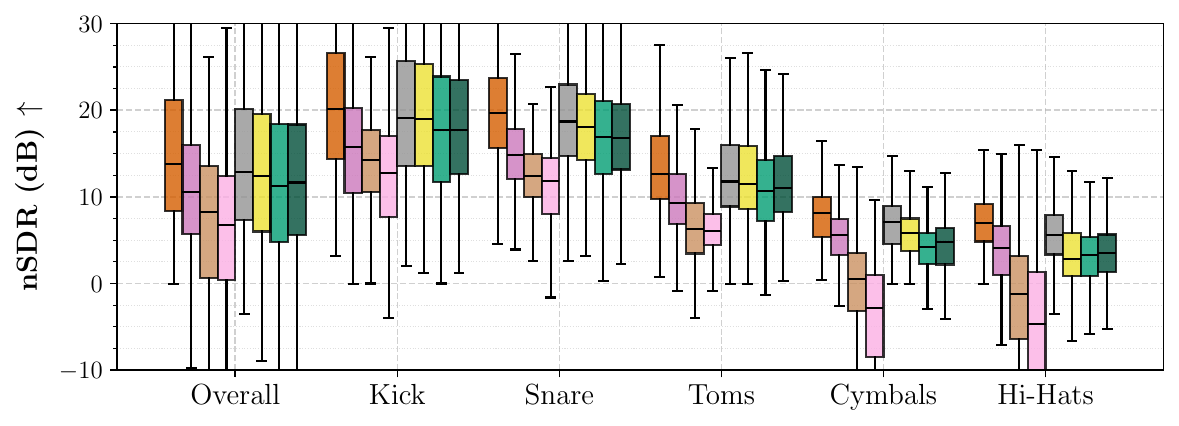}
            \label{fig:si_sdr_masked_train}
        } \\[-0.5mm]
        \subfloat{
            \includegraphics[trim={0 4pt 0 6pt}, clip, width=\textwidth]{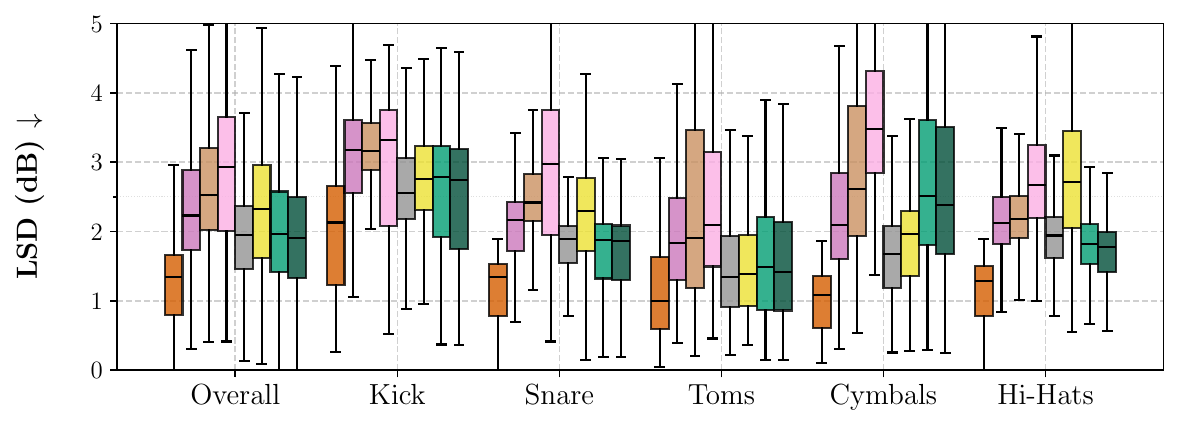}
            \label{fig:lsd_masked_train}
        } \\[-0.5mm]
        \subfloat{
            \includegraphics[trim={0 5pt 0 6pt}, clip, width=\textwidth]{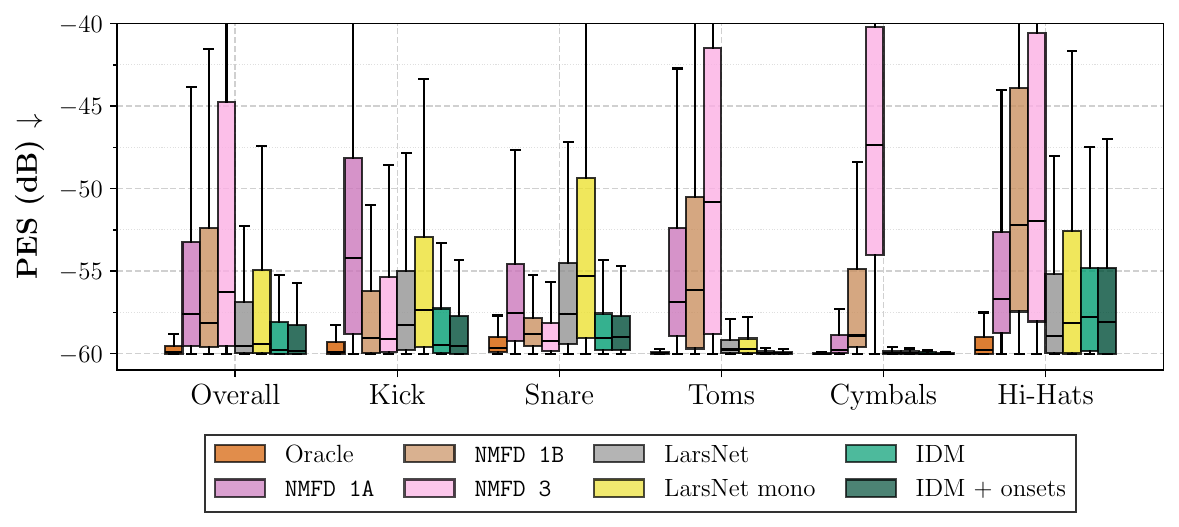}
            \label{fig:pes_masked_train}
        }
    \end{minipage}
    \hfill
    \begin{minipage}[t]{\RightColWidth}
        \centering
        \small{\quad \quad\textbf{Test kits}}
        \coltitlespace
        \subfloat{
            \includegraphics[ trim={0 4pt 0 7pt}, clip, width=\textwidth]{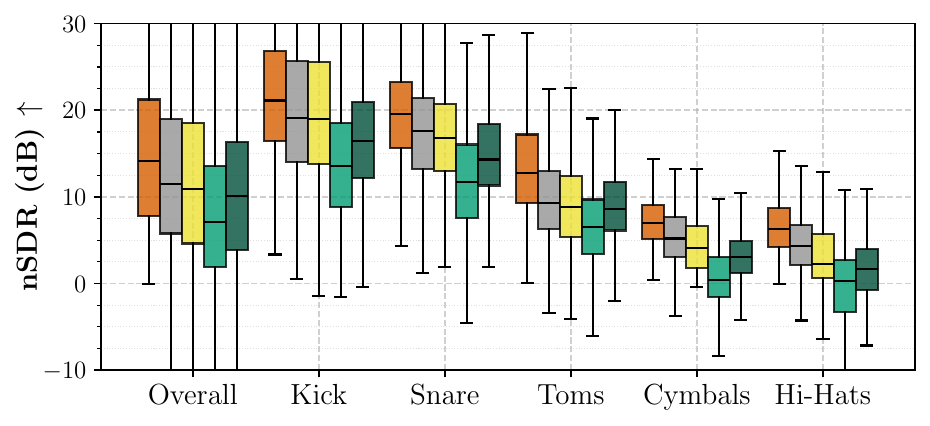}
            \label{fig:si_sdr_masked_test}
        } \\[-0.5mm]
        \subfloat{
            \includegraphics[trim={0 4pt 0 3pt}, clip, width=\textwidth]{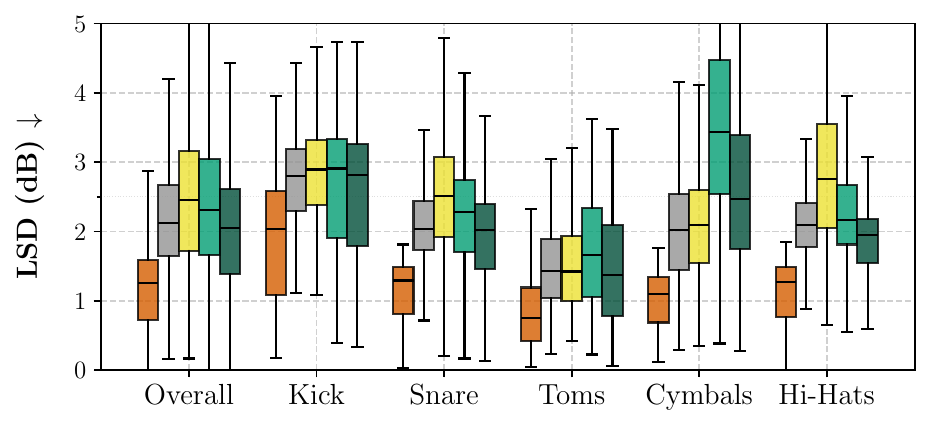}
            \label{fig:lsd_masked_test}
        } \\[-0.5mm]
        \subfloat{
            \includegraphics[trim={0 5pt 0 3pt}, clip, width=\textwidth]{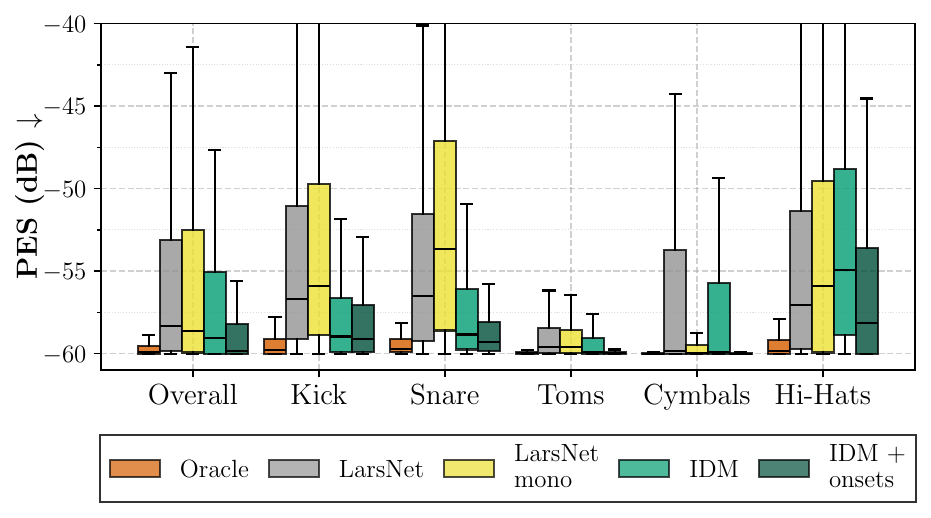}
            \label{fig:pes_masked_test}
        }
    \end{minipage}
    \caption{Comparison of masking-based separation performance metrics on the full test set of StemGMD. The left column shows results on drum kits used during training and the right column shows results on test kits. From top to bottom, the plots display \acl{nSDR} (higher is better), \acl{LSD} (lower is better), and \acl{PES} (lower is better) for masked outputs.}
    \label{fig:masked_results}
\end{figure*}

As shown in Figure~\ref{fig:transcription_results}, \new{the trained transcription module} achieves high precision and recall values (both in the $90$s percentile range) across most instruments in the test tracks. This strong performance is partly due to the controlled experimental conditions, \new{including low data diversity, as the degraded precision on the held-out kits shows}.

\sout{Nevertheless, our model significantly outperforms \nmfd~baseline approaches, which have been shown to outperform unsupervised deep learning methods in previous work. These results also highlight a disadvantage of \ac{NMFD}-based methods: even when provided with ground-truth transcription at the algorithm's initialization, the basis optimizations lead to smoothed activations that compromise transcription accuracy.}

The results reveal some variations across drum instruments. We observe particularly challenging detection patterns for \new{cymbals, and toms}. Despite these challenges, the overall transcription accuracy remains robust, providing a solid foundation for the subsequent separation task. 
    
    \subsection{Separation Results}\label{sec:separation_results}

    \subsubsection{\textbf{Direct synthesis results}}

            

Figure~\ref{fig:synth_results_final} presents the synthesis-based separation results using \acf{LSD} and silence prediction (\ac{PES}) metrics. \ours~demonstrates superior silence prediction capabilities compared to supervised baselines and achieves \sout{better} comparable \ac{LSD} scores to \larsnet~for most instruments except toms \sout{and cymbals}. Direct synthesis produces lower \ac{LSD} values than masked outputs for all classes except \new{cymbals}, suggesting our generative approach can produce spectrally accurate reconstructions without the artifacts introduced by masking.

\smallskip

    \subsubsection{\textbf{Masking-based separation}}

The results for masking-based separation are presented in Figure~\ref{fig:masked_results}. The proposed \ours~model outperforms all \nmfd~baselines by a substantial margin across all metrics, despite having access to transcription information during inference. Compared to the supervised \larsnet, our approach performs slightly worse on masked metrics, though we \sout{marginally} outperform the \larsnetmono~configuration on \ac{LSD}.

According to the \ac{nSDR} scores, a notable challenge appears in separating \new{C}ymbal\new{s}  and Hi-Hats, where the fixed one-shot duration of $1$ second in our approach might be insufficient to model their longer decay characteristics \cite{vandeveireSigmoidalNMFDConvolutional2021}, negatively affecting objective metrics without necessarily reflecting perceptual quality. 

\smallskip

\subsubsection{\textbf{Influence of onsets during inference}}

Including ground-truth onsets during inference (\oursonsets) consistently improves separation performance, occasionally surpassing \larsnetmono~or \larsnet~results (e.g., on kick for masked \ac{LSD} and Hi-Hats for synth \ac{LSD}). \new{This is very noticeable for Toms in the train kits and for Hi-Hats and Cymbals for the test kits.}. This confirms that our method's optimal performance depends on accurate onset detection, though the relatively small improvement margin reflects our model's already strong transcription capabilities.

\new{
\subsubsection{\textbf{Test kits}}
Even though \texttt{IDM}  has only been trained to synthesize $6$ drum kits, it still produces reasonable results when prompted to separate the $4$ held-out test kits. While there's significant spectral overlap between the different drum instruments, different samples from the same class share high spectral similarity and thus masking-based separation (Figure \ref{fig:masked_results}, right) and synthesis-based spectral distance (Figure \ref{fig:synth_results_final}, right) yield comparable, and sometimes better results than the strongest baselines.
}

\smallskip
\subsubsection{\textbf{Results in 9-class configuration}}

We report model results in the full \texttt{9-Class} configuration in Table~\ref{tab:metrics_comparison} for reference. The \texttt{9-Class} configuration provides a more detailed view of the model's performance across all drum instruments. Particularly notable is the disparity between open and closed hi-hats, which are often grouped in evaluation protocols such as that used by \cite{mezzaDeepDrumSource2024}.

\section{Discussion} \label{sec:discussion}

Our experimental results demonstrate that the proposed \ac{IDM} approach successfully integrates transcription and synthesis for \acl{DSS}. Without requiring access to ground-truth isolated stems during training, our model achieves performances comparable to the ones of supervised approaches that depend on such data while using $\approx 100$ times less parameters (\nparams~ vs $49.1$M). \new{In addition, while datasets with clean, multitrack drum stems are scarce and challenging to produce, numerous datasets suitable for \ac{ADT} are available, ranging from large-scale MIDI-aligned collections to smaller, manually annotated corpora of real-world recordings, making our method scalable.} The ability to directly synthesize one-shot samples for each instrument represents an additional advantage, facilitating potential applications in music production and remixing scenarios.

The transcription component can be improved by leveraging more data, state-of-the-art techniques\cite{weberRealtimeAutomaticDrum2024}, data augmentation \cite{callenderImprovingPerceptualQuality2020, zehrenHighqualityReproducibleAutomatic2023}, incorporating beat information \cite{voglDrumTranscriptionJoint2017}, and increasing the temporal resolution of onset detection. External \ac{ADT} models or even human input could be integrated to enhance separation results in practical applications.

\begin{table}
\centering
\caption{Performance metrics by instrument class}
\begin{tabular}{lccccc}
\toprule
\multirow{2}{*}{Class} & \multicolumn{3}{c}{Masked} & \multicolumn{2}{c}{Synthesized} \\
\cmidrule(lr){2-4} \cmidrule(lr){5-6}
 & nSDR$\uparrow$ & LSD$\downarrow$ & PES$\downarrow$ & LSD$\downarrow$ & PES$\downarrow$ \\
\midrule
kick & $17.47$ & $2.58$ & $-54.37$ & $1.84$ & $-55.46$ \\
snare & $17.29$ & $1.73$ & $-57.20$ & $1.83$ & $-58.87$ \\
hihat\_closed & $1.66$ & $1.90$ & $-55.88$ & $1.94$ & $-56.83$ \\
hihat\_open & $6.68$ & $1.71$ & $-59.81$ & $1.86$ & $-59.99$ \\
hi\_tom & $8.50$ & $1.16$ & $-59.57$ & $1.03$ & $-59.85$ \\
mid\_tom & $6.41$ & $1.11$ & $-59.62$ & $0.90$ & $-59.70$ \\
low\_tom & $7.35$ & $1.62$ & $-59.30$ & $1.32$ & $-59.57$ \\
ride & $2.12$ & $2.09$ & $-59.57$ & $2.24$ & $-59.65$ \\
crash\_left & $4.78$ & $2.84$ & $-59.88$ & $2.99$ & $-59.94$ \\
\bottomrule
\end{tabular}
\label{tab:metrics_comparison}
\end{table}

Our evaluation protocol reveals limitations in traditional \ac{DSS} assessment methods. Waveform metrics exhibit considerable variance across instrument classes, suggesting they may not optimally measure objective performance. The \texttt{9-class} results indicate significant performance disparities that challenge conventional class groupings, as exemplified by open hi-hats showing greater acoustic similarity to ride cymbals than to closed hi-hats. Finally, the sparsity of less frequently played instruments can bias track-level metrics due to the predominance of silence. 

This approach could benefit musical traditions with significant percussive components, such as African or Latin American music\cite{maiaNovelDatabaseBrazilian2018}, particularly in low-resource contexts where isolated stem data is scarce or unavailable.

\section{Conclusion}
\label{sec:conclusion}

This paper \new{presents} the \acf{IDM}, a novel approach to \acf{DSS} that integrates \acl{ADT} and \acl{OSS} in an end-to-end framework. By leveraging an analysis-by-synthesis methodology, our approach achieves high-quality DSS by \new{ replacing the need for isolated stems with more easily obtainable transcription data}.

Experimental results \new{demonstrate} that \ac{IDM} performs comparably to state-of-the-art supervised methods that require multitrack training data, while significantly outperforming matrix decomposition baselines. The modular synthesis framework  and the ability to directly synthesize one-shot samples for each drum instrument provide additional flexibility for creative applications.

Future work may explore extending this approach to \new{more diverse datasets} and enhancing synthesis quality. The successful conditional synthesis without exposure to ground-truth isolated targets enables potential applications with more diverse drum kits and real-world drum mixture data. Future implementations could replace the one-hot representation of drum kits with pre-trained embedding models or learned representations. \ac{IDM} could also potentially operate without transcription information by leveraging unsupervised learning approaches \cite{choiDeepUnsupervisedDrum2019}, while synthesis quality could be enhanced through adversarial learning techniques \cite{nistalDRUMGANSynthesisDrum2020}.

\section*{Acknowledgments}

This work was funded by the European Union (ERC, HI-Audio, 101052978). Views and opinions expressed are however those of the author(s) only and do not necessarily reflect those of the European Union or the European Research Council. Neither the European Union nor the granting authority can be held responsible for them.

\bibliographystyle{IEEEtran}
\bibliography{macros, references}

\begin{thebibliography}{10}
\providecommand{\url}[1]{#1}
\csname url@samestyle\endcsname
\providecommand{\newblock}{\relax}
\providecommand{\bibinfo}[2]{#2}
\providecommand{\BIBentrySTDinterwordspacing}{\spaceskip=0pt\relax}
\providecommand{\BIBentryALTinterwordstretchfactor}{4}
\providecommand{\BIBentryALTinterwordspacing}{\spaceskip=\fontdimen2\font plus
\BIBentryALTinterwordstretchfactor\fontdimen3\font minus \fontdimen4\font\relax}
\providecommand{\BIBforeignlanguage}[2]{{%
\expandafter\ifx\csname l@#1\endcsname\relax
\typeout{** WARNING: IEEEtran.bst: No hyphenation pattern has been}%
\typeout{** loaded for the language `#1'. Using the pattern for}%
\typeout{** the default language instead.}%
\else
\language=\csname l@#1\endcsname
\fi
#2}}
\providecommand{\BIBdecl}{\relax}
\BIBdecl

\bibitem{fitzgeraldAutomaticDrumTranscription2004}
D.~Fitzgerald, ``Automatic drum transcription and source separation,'' Ph.D. dissertation, Dublin Institute of Technology, Dublin, Ireland, 2004.

\bibitem{dittmarRealtimeTranscriptionSeparation2014}
C.~Dittmar and D.~G{\"a}rtner, ``Real-time transcription and separation of drum recordings based on {{NMF}} decomposition,'' in \emph{Proc. Int. Conf. Digit. Audio Effects}, 2014, pp. 187--194.

\bibitem{gilletENSTDrumsExtensiveAudiovisual2006}
O.~Gillet and G.~Richard, ``{{ENST-Drums}}: An extensive audio-visual database for drum signals processing,'' in \emph{Proc. Int. Soc. Music Inf. Retrieval Conf.}, 2006, pp. 156--159.

\bibitem{gilletTranscriptionSeparationDrum2008}
------, ``Transcription and separation of drum signals from polyphonic music,'' \emph{IEEE/ACM Trans. Audio Speech Lang. Process.}, vol.~16, no.~3, pp. 529--540, 2008.

\bibitem{dittmarReverseEngineeringAmen2016}
C.~Dittmar and M.~M{\"u}ller, ``Reverse engineering the amen break - score-informed separation and restoration applied to drum recordings,'' \emph{IEEE/ACM Trans. Audio Speech Lang. Process.}, vol.~24, no.~9, pp. 1535--1547, 2016.

\bibitem{mezzaDeepDrumSource2024}
A.~I. Mezza, R.~Giampiccolo, A.~Bernardini, and A.~Sarti, ``Toward deep drum source separation,'' \emph{Pattern Recognition Letters}, vol. 183, pp. 86--91, 2024.

\bibitem{rossingSciencePercussionInstruments2000}
T.~D. Rossing, \emph{Science of Percussion Instruments}.\hskip 1em plus 0.5em minus 0.4em\relax Singapore: World Scientific, 2000, vol.~3.

\bibitem{hennequinSpleeterFastEfficient2020}
R.~Hennequin, A.~Khlif, F.~Voituret, and M.~Moussallam, ``Spleeter: A fast and efficient music source separation tool with pre-trained models,'' \emph{J. Open Source Softw.}, vol.~5, no.~56, p. 2154, 2020.

\bibitem{defossezHybridSpectrogramWaveform2021}
A.~D{\'e}fossez, ``Hybrid spectrogram and waveform source separation,'' in \emph{Proc. ISMIR Workshop Music Source Separation}, 2021.

\bibitem{mezzaBenchmarkingMusicDemixing2024}
A.~I. Mezza, R.~Giampiccolo, A.~Bernardini, and A.~Sarti, ``Benchmarking music demixing models for deep drum source separation,'' in \emph{Proc. IEEE Int. Symp. Internet Sounds}.\hskip 1em plus 0.5em minus 0.4em\relax Erlangen, Germany: IEEE, 2024, pp. 1--6.

\bibitem{manilowSimultaneousSeparationTranscription2020a}
E.~Manilow, P.~Seetharaman, and B.~Pardo, ``Simultaneous separation and transcription of mixtures with multiple polyphonic and percussive instruments,'' in \emph{Proc. IEEE Int. Conf. Acoust. Speech Signal Process.}\hskip 1em plus 0.5em minus 0.4em\relax Barcelona, Spain: IEEE, 2020, pp. 771--775.

\bibitem{cheukEffectSpectrogramReconstruction2020}
K.~W. Cheuk, Y.-J. Luo, E.~Benetos, and D.~Herremans, ``The effect of spectrogram reconstruction on automatic music transcription: {{An}} alternative approach to improve transcription accuracy,'' in \emph{Proc. Int. Conf. Pattern Recognit.}\hskip 1em plus 0.5em minus 0.4em\relax Milan, Italy: IEEE, 2020, pp. 9091--9098.

\bibitem{choiDeepUnsupervisedDrum2019}
K.~Choi and K.~Cho, ``Deep unsupervised drum transcription,'' in \emph{Proc. Int. Soc. Music Inf. Retrieval Conf.}, Delft, Netherlands, 2019, pp. 183--191.

\bibitem{schulze-forsterUnsupervisedMusicSource2023}
K.~{Schulze-Forster}, G.~Richard, L.~Kelley, C.~S. Doire, and R.~Badeau, ``Unsupervised music source separation using differentiable parametric source models,'' \emph{IEEE/ACM Trans. Audio Speech Lang. Process.}, vol.~31, pp. 1276--1289, 2023.

\bibitem{vandeveireSigmoidalNMFDConvolutional2021}
L.~Vande~Veire, C.~De~Boom, and T.~De~Bie, ``Sigmoidal {{NMFD}}: Convolutional {{NMF}} with saturating activations for drum mixture decomposition,'' \emph{Electronics}, vol.~10, no.~3, p. 284, 2021.

\bibitem{shierDifferentiableModellingPercussive2023}
J.~Shier, F.~Caspe, A.~Robertson, M.~Sandler, C.~Saitis, and A.~McPherson, ``Differentiable modelling of percussive audio with transient and spectral synthesis,'' in \emph{Proc. Forum Acusticum}, 2023.

\bibitem{wuReviewAutomaticDrum2018}
C.-W. Wu, C.~Dittmar, C.~Southall, R.~Vogl, G.~Widmer, J.~Hockman, M.~M{\"u}ller, and A.~Lerch, ``A review of automatic drum transcription,'' \emph{IEEE/ACM Trans. Audio Speech Lang. Process.}, vol.~26, no.~9, pp. 1457--1483, 2018.

\bibitem{lindsay-smithDrumkitTranscriptionConvolutive2012}
H.~{Lindsay-Smith}, S.~McDonald, and M.~Sandler, ``Drumkit transcription via convolutive {{NMF}},'' in \emph{Proc. Int. Conf. Digit. Audio Effects}, York, UK, 2012.

\bibitem{voglDrumTranscriptionJoint2017}
R.~Vogl, M.~Dorfer, G.~Widmer, and P.~Knees, ``Drum transcription via joint beat and drum modeling using convolutional recurrent neural networks,'' in \emph{Proc. Int. Soc. Music Inf. Retrieval Conf.}, Suzhou, China, 2017, pp. 150--157.

\bibitem{callenderImprovingPerceptualQuality2020}
L.~Callender, C.~Hawthorne, and J.~Engel, ``Improving perceptual quality of drum transcription with the expanded groove {{MIDI}} dataset,'' \emph{arXiv preprint arXiv:2004.00188}, 2020.

\bibitem{voglMultiinstrumentDrumTranscription2018}
R.~Vogl, G.~Widmer, and P.~Knees, ``Towards multi-instrument drum transcription,'' in \emph{Proc. Int. Conf. Digit. Audio Effects}, Aveiro, Portugal, 2018, pp. 57--64.

\bibitem{zehrenHighqualityReproducibleAutomatic2023}
M.~Zehren, M.~Alunno, and P.~Bientinesi, ``High-quality and reproducible automatic drum transcription from crowdsourced data,'' \emph{Signals}, vol.~4, no.~4, pp. 768--787, 2023.

\bibitem{weberRealtimeAutomaticDrum2024}
P.~Weber, C.~Uhle, M.~M{\"u}ller, and M.~Lang, ``Real-time automatic drum transcription using dynamic few-shot learning,'' in \emph{Proc. IEEE Int. Symp. Internet Sounds}.\hskip 1em plus 0.5em minus 0.4em\relax Erlangen, Germany: IEEE, 2024, pp. 1--8.

\bibitem{cookPhysicallyInformedSonic1996}
P.~R. Cook, ``Physically informed sonic modeling ({{PhISM}}): {{Percussive}} synthesis,'' in \emph{Proc. Int. Comput. Music Conf.}, 1996, pp. 228--231.

\bibitem{hanWAV2SHAPEHearingShape2020}
H.~Han and V.~Lostanlen, ``Wav2shape: {{Hearing}} the shape of a drum machine,'' in \emph{Proc. Forum Acusticum}, 2020, pp. 647--654.

\bibitem{iiiPARSHLAnalysisSynthesis1987}
J.~O.~S. III and X.~Serra, ``{{PARSHL}}: An analysis/synthesis program for non-harmonic sounds based on a sinusoidal representation,'' in \emph{Proc. Int. Comput. Music Conf.}, 1987.

\bibitem{vermaExtendingSpectralModeling2000a}
T.~S. Verma and T.~H.-Y. Meng, ``Extending spectral modeling synthesis with transient modeling synthesis,'' vol.~24, no.~2, pp. 47--59, 2000.

\bibitem{ramiresNeuralPercussiveSynthesis2020}
A.~Ramires, P.~Chandna, X.~Favory, E.~G{\'o}mez, and X.~Serra, ``Neural {{Percussive Synthesis Parameterised}} by {{High-Level Timbral Features}},'' in \emph{Proc. IEEE Int. Conf. Acoust. Speech Signal Process.}\hskip 1em plus 0.5em minus 0.4em\relax Barcelona, Spain: IEEE, 2020, pp. 786--790.

\bibitem{aouameurNeuralDrumMachine2019}
C.~Aouameur, P.~Esling, and G.~Hadjeres, ``Neural drum machine : {{An}} interactive system for real-time synthesis of drum sounds,'' in \emph{Proc. Int. Conf. Comput. Creativity}, 2019, pp. 92--99.

\bibitem{nistalDRUMGANSynthesisDrum2020}
J.~Nistal, S.~Lattner, and G.~Richard, ``{{DRUMGAN}}: {{Synthesis}} of drum sounds with timbral feature conditioning using generative adversarial networks,'' in \emph{Proc. Int. Soc. Music Inf. Retrieval Conf.}, Utrecht, Netherlands, 2020, pp. 590--597.

\bibitem{drysdaleAdversarialSynthesisDrum2020}
J.~Drysdale, M.~Tomczak, and J.~Hockman, ``Adversarial synthesis of drum sounds,'' in \emph{Proc. Int. Conf. Digit. Audio Effects}, Vienna, Austria, 2020, pp. 167--172.

\bibitem{rouardCRASHRawAudio2021}
S.~Rouard and G.~Hadjeres, ``{{CRASH}}: Raw audio score-based generative modeling for controllable high-resolution drum sound synthesis,'' in \emph{Proc. Int. Soc. Music Inf. Retrieval Conf.}, 2021, pp. 579--585.

\bibitem{simionatoSinesTransientNoise2025}
R.~Simionato and S.~Fasciani, ``Sines, transient, noise neural modeling of piano notes,'' \emph{Frontiers in Signal Processing}, vol.~4, p. 1494864, 2025.

\bibitem{smaragdisNonnegativeMatrixFactor2004}
P.~Smaragdis, ``Non-negative matrix factor deconvolution; extraction of multiple sound sources from monophonic inputs,'' in \emph{Proc. Intl. Conf. on Independent Component Analysis and Blind Signal Separation}.\hskip 1em plus 0.5em minus 0.4em\relax Grenada, Spain: Springer, 2004, pp. 494--499.

\bibitem{caiDualchannelDrumSeparation2021}
C.-Y. Cai, Y.-H. Su, and L.~Su, ``Dual-channel drum separation for low-cost drum recording using non-negative matrix factorization,'' in \emph{Proc. Asia-Pacific Signal Inf. Process. Assoc. Annu. Summit Conf.}\hskip 1em plus 0.5em minus 0.4em\relax Tokyo, Japan: IEEE, 2021, pp. 17--22.

\bibitem{gillickLearningGrooveInverse2019}
J.~Gillick, A.~Roberts, J.~Engel, D.~Eck, and D.~Bamman, ``Learning to groove with inverse sequence transformations,'' in \emph{Proc. Int. Conf. Mach. Learning}, Long Beach, USA, 2019, pp. 2269--2279.

\bibitem{engelDDSPDifferentiableDigital2020}
J.~H. Engel, L.~Hantrakul, C.~Gu, and A.~Roberts, ``{{DDSP}}: {{Differentiable}} digital signal processing,'' in \emph{Proc. Int. Conf. Learn. Representations}, 2020.

\bibitem{masudaSynthesizerSoundMatching2021}
N.~Masuda and D.~Saito, ``Synthesizer sound matching with differentiable {{DSP}}.'' in \emph{Proc. Int. Soc. Music Inf. Retrieval Conf.}, 2021, pp. 428--434.

\bibitem{torresUnsupervisedHarmonicParameter2024}
B.~Torres, G.~Peeters, and G.~Richard, ``Unsupervised harmonic parameter estimation using differentiable {{DSP}} and spectral optimal transport,'' in \emph{Proc. IEEE Int. Conf. Acoust. Speech Signal Process.}\hskip 1em plus 0.5em minus 0.4em\relax Seoul, South Korea: IEEE, 2024, pp. 1176--1180.

\bibitem{hayesReviewDifferentiableDigital2024}
B.~Hayes, J.~Shier, G.~Fazekas, A.~McPherson, and C.~Saitis, ``A review of differentiable digital signal processing for music and speech synthesis,'' \emph{Frontiers in Signal Processing}, vol.~3, p. 1284100, 2024.

\bibitem{kawamuraDifferentiableDigitalSignal2022}
M.~Kawamura, T.~Nakamura, D.~Kitamura, H.~Saruwatari, Y.~Takahashi, and K.~Kondo, ``Differentiable digital signal processing mixture model for synthesis parameter extraction from mixture of harmonic sounds,'' in \emph{Proc. IEEE Int. Conf. Acoust. Speech Signal Process.}\hskip 1em plus 0.5em minus 0.4em\relax Singapore, Singapore: IEEE, 2022, pp. 941--945.

\bibitem{richardFullyDifferentiableModel2024}
G.~Richard, P.~Chouteau, and B.~Torres, ``A fully differentiable model for unsupervised singing voice separation,'' in \emph{Proc. IEEE Int. Conf. Acoust. Speech Signal Process.}\hskip 1em plus 0.5em minus 0.4em\relax Seoul, South Korea: IEEE, 2024, pp. 946--950.

\bibitem{liuConvNet2020s2022}
Z.~Liu, H.~Mao, C.-Y. Wu, C.~Feichtenhofer, T.~Darrell, and S.~Xie, ``A {{ConvNet}} for the 2020s,'' in \emph{Proc. IEEE/CVF Conf. Comput. Vis. Pattern Recognit.}\hskip 1em plus 0.5em minus 0.4em\relax New Orleans, USA: IEEE, 2022, pp. 11\,966--11\,976.

\bibitem{hawthorneOnsetsFramesDualobjective2018}
C.~Hawthorne, E.~Elsen, J.~Song, A.~Roberts, I.~Simon, C.~Raffel, J.~H. Engel, S.~Oore, and D.~Eck, ``Onsets and frames: {{Dual-objective}} piano transcription,'' in \emph{Proc. Int. Soc. Music Inf. Retrieval Conf.}, E.~G{\'o}mez, X.~Hu, E.~Humphrey, and E.~Benetos, Eds., Paris, France, 2018, pp. 50--57.

\bibitem{perezFiLMVisualReasoning2017}
E.~Perez, F.~Strub, H.~{de Vries}, V.~Dumoulin, and A.~Courville, ``{{FiLM}}: {{Visual Reasoning}} with a {{General Conditioning Layer}},'' \emph{arXiv preprint arXiv:1709.07871}, no. arXiv:1709.07871, 2017.

\bibitem{liutkusGeneralizedWienerFiltering2015}
A.~Liutkus and R.~Badeau, ``Generalized {{Wiener}} filtering with fractional power spectrograms,'' in \emph{Proc. IEEE Int. Conf. Acoust. Speech Signal Process.}\hskip 1em plus 0.5em minus 0.4em\relax Brisbane, Australia: IEEE, 2015, pp. 266--270.

\bibitem{baLayerNormalization2016}
J.~L. Ba, J.~R. Kiros, and G.~E. Hinton, ``Layer normalization,'' \emph{arXiv preprint arXiv:1607.06450}, 2016.

\bibitem{bockEvaluatingOnlineCapabilities2012}
S.~B{\"o}ck, F.~Krebs, and M.~Schedl, ``Evaluating the online capabilities of onset detection methods,'' in \emph{Proc. Int. Soc. Music Inf. Retrieval Conf.}, Porto, Portugal, 2012, pp. 49--54.

\bibitem{wangNeuralSourcefilterWaveform2019}
X.~Wang, S.~Takaki, and J.~Yamagishi, ``Neural source-filter waveform models for statistical parametric speech synthesis,'' \emph{IEEE/ACM Trans. Audio Speech Lang. Process.}, vol.~28, pp. 402--415, 2019.

\bibitem{lopez-serranoNMFToolboxMusic2019}
P.~{L{\'o}pez-Serrano}, C.~Dittmar, Y.~{\"O}zer, and M.~M{\"u}ller, ``{{NMF}} toolbox: {{Music}} processing applications of nonnegative matrix factorization,'' in \emph{Proc. Int. Conf. Digit. Audio Effects}, vol.~19, 2019, pp. 2--6.

\bibitem{raffelMIR_EVALTransparentImplementation2014}
C.~Raffel, B.~McFee, E.~J. Humphrey, J.~Salamon, O.~Nieto, D.~Liang, and D.~P.~W. Ellis, ``{{MIR}}\_{{EVAL}}: {{A}} transparent implementation of common {{MIR}} metrics,'' in \emph{Proc. Int. Soc. Music Inf. Retrieval Conf.}, Taipei, Taiwan, 2014, pp. 367--372.

\bibitem{mitsufujiMusicDemixingChallenge2021}
Y.~Mitsufuji, G.~Fabbro, S.~Uhlich, and F.-R. St{\"o}ter, ``Music demixing challenge2021,'' \emph{arXiv preprint arXiv:2108.13559}, 2021.

\bibitem{torcoliObjectiveMeasuresPerceptual2021}
M.~Torcoli, T.~Kastner, and J.~Herre, ``Objective {{Measures}} of {{Perceptual Audio Quality Reviewed}}: {{An Evaluation}} of {{Their Application Domain Dependence}},'' \emph{IEEE/ACM Trans. Audio Speech Lang. Process.}, vol.~29, pp. 1530--1541, 2021.

\bibitem{schulze-forsterWeaklyInformedAudio2019}
K.~{Schulze-Forster}, C.~S.~J. Doire, G.~Richard, and R.~Badeau, ``Weakly informed audio source separation,'' in \emph{Proc. IEEE Workshop Appl. Signal Process. Audio Acoust.}\hskip 1em plus 0.5em minus 0.4em\relax IEEE, 2019, pp. 273--277.

\bibitem{maiaNovelDatabaseBrazilian2018}
L.~Maia, P.~D. {de Tomaz Jr.}, M.~Fuentes, M.~Rocamora, L.~W.~P. Biscainho, M.~V.~M. {da Costa}, and S.~Cohen, ``A novel database of brazilian rhythmic instruments and some experiments in computational rhythm analysis,'' in \emph{Proc. Audio Eng. Soc. Latin Amer. Conf.}, Montevideo, Uruguay, 2018.

\end{thebibliography}

\newpage

 




\vfill

\end{document}